\documentclass[preprint2]{aastex}
\usepackage{amsmath,latexsym}
\usepackage{multirow}

\newcommand{\myemail}{siegfried.eggl@univie.ac.at}

\shorttitle{Detectability of Earth-like Planets in Circumstellar Habitable Zones of 
Binary Star Systems with Sun-like Components}
\shortauthors{Eggl et al.}

\begin{document}

\title{Detectability of Earth-like Planets in Circumstellar Habitable Zones of Binary 
Star Systems with Sun-like Components}

\author{Siegfried Eggl\altaffilmark{1}, Nader Haghighipour\altaffilmark{2},  
Elke Pilat-Lohinger\altaffilmark{1}}
\email{\myemail}

 \altaffiltext{1}{University of Vienna, IfA, T\"urkenschanzstr. 17, A-1180 Vienna, Austria}

 \altaffiltext{2}{Institute for Astronomy and NASA Astrobiology Institute, 2680 Woodlawn Drive, 
Honolulu, HI 96822, USA}

\begin{abstract}
Given the considerable percentage of stars that are members of binaries or stellar multiples in the 
Solar neighborhood, it is expected that many of these binaries host planets, possibly even habitable ones.
The discovery of a terrestrial planet in the $\alpha$ Centauri system supports this notion.
Due to the potentially strong gravitational interaction that an Earth-like planet may experience in such 
systems, classical approaches to determining habitable zones, especially in close S-Type binary systems, 
can be rather inaccurate. Recent progress in this field, however, allows to identify regions around the 
star permitting permanent habitability. 
While the discovery of $\alpha$ Cen Bb has shown that terrestrial planets can be detected in solar-type binary stars using current observational facilities, it
remains to be shown whether this is also the case for Earth analogues in habitable zones. 
We provide analytical expressions for the maximum and RMS values of radial velocity
and astrometric signals, as well as transit probabilities of terrestrial planets in
such systems, showing that the dynamical interaction of the second star with
the planet may indeed facilitate the planet’s detection.
As an example, we discuss the detectability of additional Earth-like planets in the averaged, extended, and 
permanent habitable zones around both stars of the $\alpha$ Centauri system. 
\end{abstract}


\section{Introduction}

The past decades have seen a great number of discoveries of planets around stars other than our Sun 
\citep{schneider-2011}. As some of these planets are of terrestrial nature, hopes to identifying Earth 
analogues have lead to considerable advances towards the detection of possibly habitable worlds 
\citep{kepler2-2011,ford-et-al-2012}. Even though quite frequent in the Solar neighborhood 
\citep{kiseleva-eggleton-eggleton-2001}, not many attempts have yet been made to specifically target binary 
stars in this endeavor. Nonetheless, more than 60 planets have already been found in and around such systems 
\citep{haghighipour-2010,doyle-et-al-2011,welsh-et-al-2012,roell-et-al-2012,
orosz-et-al-2012a,orosz-et-al-2012b,dumusque2012}. Although several P-Type (circumbinary) planets 
orbiting both stars of a close binary have also been found 
\citep{doyle-et-al-2011,welsh-et-al-2012,orosz-et-al-2012a,orosz-et-al-2012b},
most planets are in the so-called S-Type \citep{rabl-dvorak-1988} 
configuration where the planet orbits only one of the binary's stars.  
A prominent example of an S-Type system is $\alpha$ Centauri AB which hosts a terrestrial planet around the fainter binary component, $\alpha$ Cen B \citep{dumusque2012}.  

The reason for the general reluctance to include binary systems in the search for 
terrestrial, habitable planets lies in the assumption that the additional interactions 
with a massive companion will make planets harder to find. That is primarily because the gravitational 
interaction between the second star and a planet may alter the planet's orbit significantly and complicate 
the task of interpreting the planetary signal. One aim of this work is, therefore, to show that changes in the planet's orbit can actually enhance 
its detectability (see section \ref{sec:results}).
Of course, the orbit of a binary as well as its stellar parameters have to be well determined in order to be able to identify signals 
from additional terrestrial planets. Sensing the need for a better understanding of binary star systems, 
efforts have been intensified to improve physical as well as orbital data for nearby binaries 
\citep[e.g.][]{torres-et-al-2010} and to evolve existing data analysis methodologies 
\citep{chauvin-et-al-2011,haghighipour-2010,pourbaix-2002,pourbaix-et-al-2002}. 

Understanding the complex interactions between a stellar binary and a planet is essential if 
a system's potential habitability is to be evaluated.
For instance, one of the main assumptions of classical habitability, as 
introduced by \citet{kasting-et-al-1993}, is that the planet moves around its host star on a circular 
orbit. This may not be a valid assumption for planets in a binary star system where the gravitational 
perturbation of the secondary can excite the eccentricity of the planet's orbit 
\citep{marchal-1990,georgakarakos-2002,eggl-et-al-2012}. \citet{eggl-et-al-2012} found
that except for S-Type systems where the secondary star is much more luminous than the planet's host star, 
variations in planetary orbit around the planet-hosting star are the main cause for changes in insolation.    
Even though \citet{eggl-et-al-2012} gave an analytic recipe for calculating the boundaries of the HZs 
in S-Type binaries, it remains to be seen whether an Earth-like planet in the HZ of a system with two 
Sun-like stars will in fact be detectable. 

In order to answer this question, we consider three techniques, namely, 
radial velocity (RV), astrometry (AM) and transit photometry (TP), and discuss whether the current 
observational facilities are capable of detecting habitable planets in such systems.
We provide analytical formula for estimating the strength of RV and AM signals for habitable, Earth-like 
planets, and show that the planet-binary interaction can enhance the chances for the detection of these
objects.

The rest of this article will be structured as follows.
In sections \ref{sec:rv} and \ref{sec:astrometry}  analytic estimates of the maximum and root mean square 
(RMS) of the strength of an RV and an AM signal that an Earth-like planet produces in an S-Type binary 
configuration will be derived. Section \ref{sec:tp} will deal with consequences of such a setup for TP. 
We will then briefly recall the different types of HZs for S-Type binaries established in \citet{eggl-et-al-2012},
and use their methodology to identify similar habitable regions in the $\alpha$ Centauri system 
(section \ref{sec:results}). This system has been chosen because first,   
it has inspired many studies on the possibility of the formation and detection of habitable planets 
around its stellar components \citep{forgan-2012,guedes-et-al-2008,thebault-et-al-2009} and second,
 \citet{dumusque2012} have already discovered an Earth-sized planet in a short-period orbit around its secondary 
star.
Therefore, we will compare our RV estimates to the actual signal of $\alpha$ Cen Bb, 
and study its influence on an additional terrestrial planet presumed in $\alpha$ Cen B's HZ. 
Finally, in section \ref{sec:dis}, the projected RV, AM and TP traces that terrestrial planets will
leave in the HZ of the $\alpha$ Centauri system are analyzed, and the results are discussed within the context 
of the sensitivity of the current observational facilities.

\section{Radial Velocity}\label{sec:rv} 

To estimate the RV signal that an Earth-like planet produces in an S-Type binary system, we will build upon the 
formalism presented by \citet{beauge-et-al-2007}.  We assume that the non-planetary contributions
 to the host star's RV signal (such as the RV variation caused by the motion
of the binary around its center of gravity) 
are known and have been subtracted, leaving behind only the residual signal due to the planet.
The motion of the planet around its host star then constitutes a perturbed two body problem, where the gravitational influence of the secondary star is still playing a role and is mirrored in the 
forced variations of the planet's orbit.

In practice, the extraction of the planetary signal is all but a trivial task. Even after subtraction of the binary's barycentric and proper motion, 
the residual will contain contributions from the binary's orbital uncertainties as well as from non-gravitational
sources which could be orders of magnitude larger than the star's reflex signal, such as the Rossiter-McLaughlin effect in transiting systems, for example \citep{ohta-et-al-2005}.  
The discovery of $\alpha$ Cen Bb showed, however, 
that a substantial reduction of non-planetary RV interference is possible if the respective binary star has been studied in sufficient detail.

The amplitude of the planet induced RV signal of the host star, $V_r$, is given by

\begin{equation}
V_r=K\; [\cos(f+\omega)+e \cos{\omega}]\,, 
\label{eq:vr}
\end{equation}

\noindent
where $K$ is equal to

\begin{equation}
K =\frac{\mu \left(\kappa \;n_p \right)^{1/3} \sin{i}}{\sqrt{1-e^2}}\,.
\label{eq:vrk}
\end{equation}

\noindent
In equation (\ref{eq:vrk}), $\mu=m_1/(m_0+m_1)$ is the planet to star mass-ratio with $m_1$ and $m_0$ being
the masses of the planet and host-star, respectively. The planet's mean motion, ${n_p}=2\pi/{P_p}$, 
is given by $n_p=\sqrt{\kappa/a^3}$ with \mbox{$\kappa = \mathcal{G} (m_0 + m_1)$},
and $P_p$ and $\mathcal{G}$ being the planet's orbital period and gravitational constant. 
The quantities $a, e, i, f$ and $\omega$ 
in equations (1) and (2) denote the planet's semimajor axis, eccentricity, orbital inclination relative 
to the plane of the sky, true-anomaly, and argument of periastron, respectively.

\begin{figure}
\center
\includegraphics[angle=-90, scale=0.40]{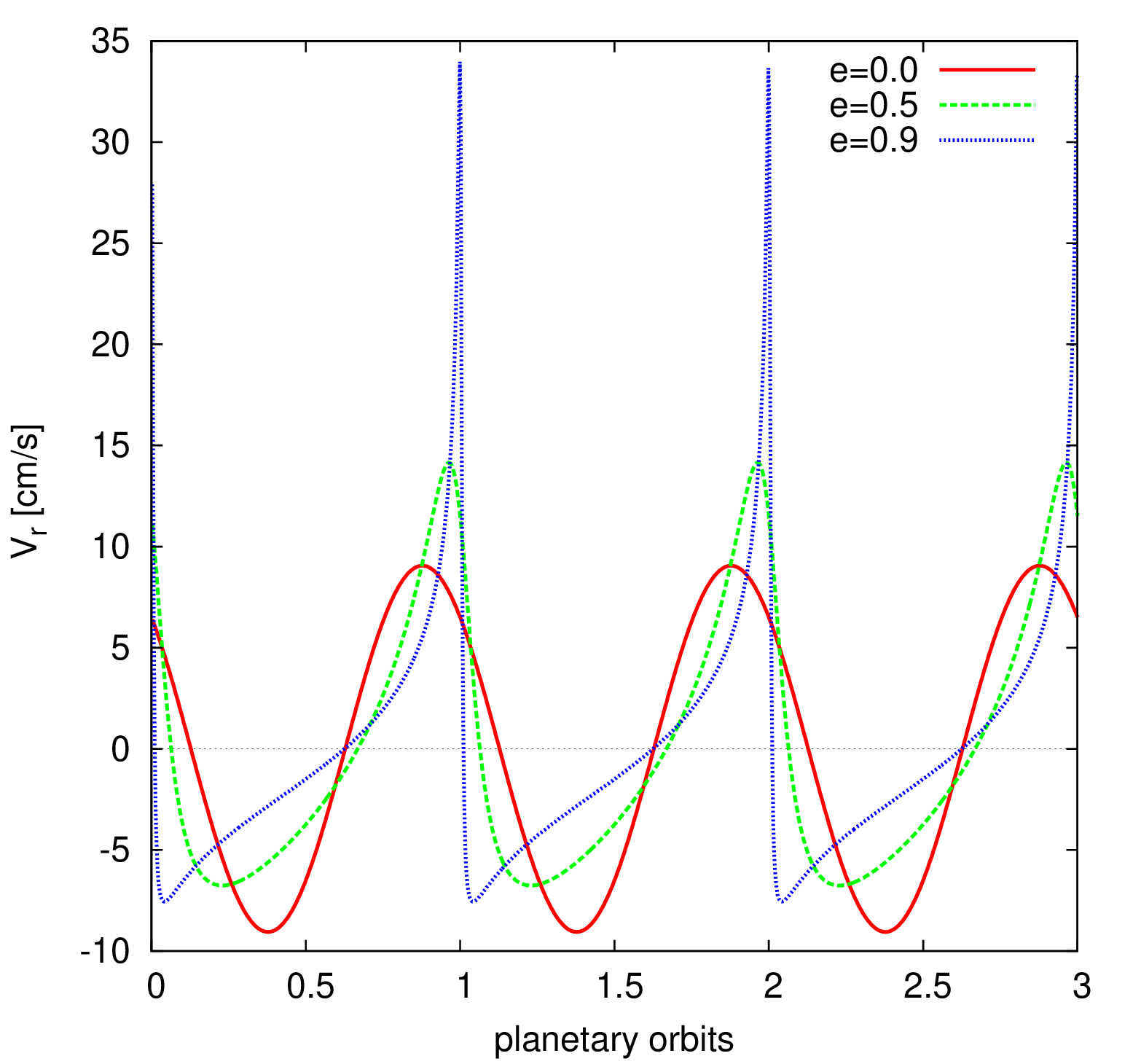}
\vskip 10pt
\includegraphics[angle=-90, scale=0.40]{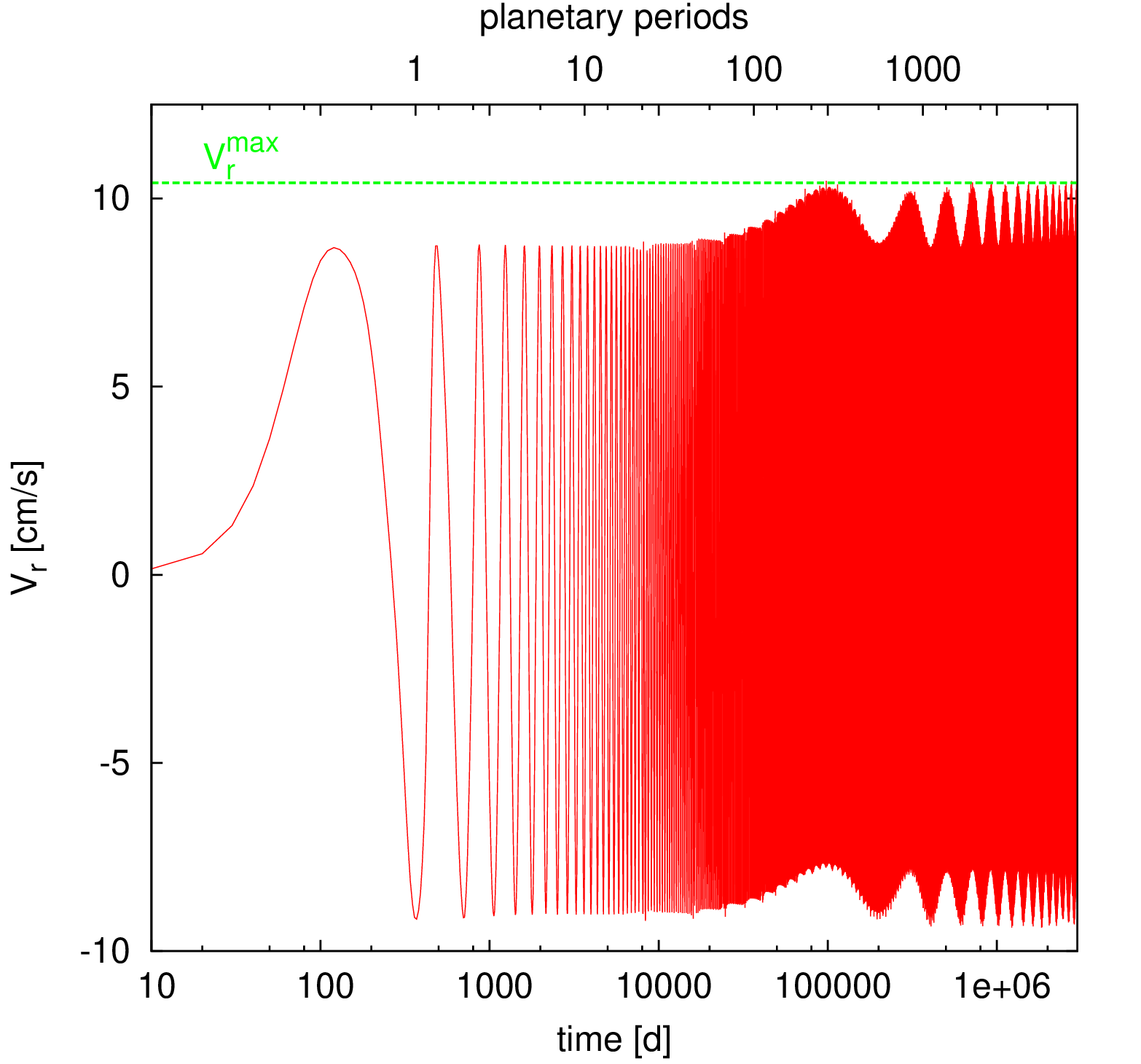}
\caption{\textit{Top:} The radial velocity signal caused by an Earth-like planet orbiting a Sun-like star with different eccentricities \citep{beauge-et-al-2007}. 
The planet is at 1 AU with $\omega=45^\circ$ when $e\ne 0$. \textit{Bottom:}
The amplitude variations of the primary's radial velocity signal due to an Earth-like planet that is subject to the gravitational perturbations of a second star. 
Both stars are Sun-like with a separation of 20 AU and an orbital eccentricity of 0.5. The planet's initial orbit was circular with a semimajor axis of 1 AU. 
Our analytically estimated maximum amplitude $V_r^{max}$ is also shown.
\label{fig1}} 
\end{figure}

Our goal in this section is to identify the range of the possible peak amplitudes that a terrestrial 
planet in an S-Type binary configuration can produce.  We note that
the gravitational influence of the second star causes the planet's orbital elements to vary, thus
inducing additional time dependent changes in the radial velocity signal $V_r$ \citep{lee-peale-2003}. 
While we know from secular perturbation theory that $a$ does not change significantly with time for hierarchical systems such
as the one under consideration \citep{marchal-1990,georgakarakos-2003}, $\omega$ becomes a function of time.
We assume coplanar orbits of the planet and the binary star which results in the planet's inclination to the plane of the sky ($i$) to remain constant. 
In contrast, the planetary eccentricity will vary between zero and a maximum $e^{max}$, where the latter value can be expressed as a function 
of the system's masses and the binary's orbital parameters \citep{eggl-et-al-2012}.
This is important, because the reflex RV signal ($V_r$) of a star can be increased significantly by planetary orbital eccentricities (Fig.~\ref{fig1}).
Using equation (\ref{eq:vr}) we identify the global maximum of $V_r$ at $f=\omega=0$, when $e=e^{max}$.
This leads to 

\begin{equation}
V_r^{max} = V_r^{circ}\, \sqrt{\frac{1+e^{max}}{1-e^{max}}}\,, \label{eq:vrmax}
\end{equation}

\noindent
where

\begin{equation}
V_r^{circ} = \frac{\sqrt{\mathcal{G}}\;m_1 \sin{i}}{\sqrt{a (m_0 + m_1)}}\,.
\end{equation}

\noindent
Equation (\ref{eq:vrmax}) presents a fully analytic estimate of the expected maximum RV signal that
a terrestrial planet produces in an S-Type binary configuration\footnote{Larger signals are possible, 
if the terrestrial planet has a considerable initial eccentricity after its formation and migration phase. 
Yet, due to the eccentricity dampening in protoplanetary discs
this seems unlikely \citep{paardekooper-leinhardt-2010}.}.

As an example for the influence of a double star on a planetary RV signal, 
the induced variations in the RV of the planet's host-star are presented in figure \ref{fig1}. 
The host-star is a constituent of a Solar-type binary with a semimajor axis of 20 AU and an orbital eccentricity of 0.5. 
Changes in the amplitude of $V_r$ are due to variations in the planet's eccentricity.
  
Since we do not know the state of the planet's orbital eccentricity at the time of observation, 
we consider a range for the maximum possible amplitudes of its radial velocity,

\begin{equation}
V_r^{circ} \leq V_r|_{f=\omega=0} \leq {V_r^{max}}\,. \label{eq:range}
\end{equation}

\noindent
% An analytic expression for $e^{max}$ can be found in \citet{eggl-et-al-2012}.
Although the range of the amplitude of the host star's RV signal, as given by equation (\ref{eq:range}),
can be used to identify the "best case" detectability limits, the maximum values of the RV signal due to the
planet will be "snapshots" that are reached only during brief moments. As a result, their values for 
assessing the precision needed to trace fingerprints of an Earth-like planet are rather limited.
In such cases, expressions for the RMS of the astrometric signal are preferable.  

Since RMS values are by convention time-averaged, we substitute $f$ by the mean anomaly $M={n_p} t$ in all corresponding functions
using the equation of the center expansion up to the sixth order in planetary eccentricities (see appendix \ref{sec:eoc}) 
and average over $M$ and $\omega$. 
The vastly different rates of change of these quantities (\mbox{$\dot{M}\gg\dot{\omega}$}) make it possible to 
consider $\omega$ to remain constant during one cycle of $M$, so that independent 
averaging can be performed. In order to eliminate short term variations in the RV signal, we first 
average over $M$. Averaging over $\omega$ as well might be desirable if for example 
the initial state of $\omega$ is unknown, or if observations stretch beyond secular evolution timescales of the 
planets argument of pericenter. We, therefore, define two different types of RMS evaluations for a 
square-integrable function $F$; 

\begin{equation}
\langle\langle F \rangle\rangle_M = \langle F^2 \rangle_M^{1/2}= 
\left[\frac{1}{2\pi}\int_0^{2\pi}F^2(M) dM \right]^{1/2},  
\label{eq:rms} 
\end{equation}

\noindent
and

\begin{equation}
\langle\langle F \rangle\rangle_{M,\omega} = 
\frac{1}{2\pi}\left[\iint\limits_0^{\quad\, 2\pi} {F^2}(M,\omega) dM d\omega \right]^{1/2}.
\label{eq:rms2}
\end{equation}

\noindent
Using equations (\ref{eq:rms}) and (\ref{eq:rms2}), the RMS values of $V_r$ are then given by

\begin{eqnarray}
\begin{split}
\langle\langle V_r \rangle\rangle_{M} =& \> \langle\langle V_r \rangle\rangle_{M,\omega}\;\times\\
&\left\{1 - 
\left[\frac{\langle e^2\rangle_M}{4} +O(\langle e^2 \rangle_M^2)\right] \cos(2 \omega)\right\}^{1/2}, 
\label{eq:vrrms}
\end{split}
\end{eqnarray}

\noindent
with

\begin{equation}
\langle\langle V_r \rangle\rangle_{M,\omega} =
\frac{\sqrt{\mathcal{G}}\;m_1|\sin{i}|}{\sqrt{2 a (m_0 + m_1)}} = \frac{1}{\sqrt{2}}V_r^{circ}.
\end{equation}

\noindent
Here we have considered $\langle a \rangle_{M} = a$ since \mbox{$\dot{a} \simeq 0$} \citep{marchal-1990,georgakarakos-2003}. Also,
 
\begin{displaymath}
\int_0^{2\pi}{{\langle e^2\rangle}_M}\>\cos(2\omega)\> d\omega=0,
\end{displaymath}

\noindent
as indicated in appendix \ref{sec:ave2}. 
It is noteworthy that the averaging over $\omega$ causes the RMS value of the RV signal to become 
independent of $e$ so that its difference with the peak signal in the circular case ($V_r^{circ}$) becomes 
a mere factor of $1/\sqrt{2}$. Thanks to their intricate relation to power-spectra, RMS values can also 
be valuable for orbit-fitting. The choice of singly or doubly averaged RMS relations for this 
purpose will depend on how many planetary orbital periods are available in the data set. 
In the case of $\alpha$ Cen Bb, there are order-of-magnitude differences in the rates of change of the mean anomaly ($\dot{M}$) and the argument of pericenter ($\dot{\omega}$). 
It would, therefore, make more sense to assume $\omega$ to be constant and add it as a variable 
in the fitting process. If stronger perturbations or additional forces act on the planet, the periods can be considerably 
shorter, so that the fully averaged equations might come in handy.

\section{Astrometry}\label{sec:astrometry} 

In order to derive the maximum and RMS values for an astrometric signal, we will use the framework presented in \citet{pourbaix-2002}. 
We again assume that the non-planetary contributions have been subtracted from the combined 
signal of the host star and planet.
The projected motion of the planet on the astrometric plane is then given by 

\begin{eqnarray}
\begin{split}
x_E &=& A\big(\cos{E} - e\big) + F \sqrt{1 - e^2}\sin{E},\\ 
y_E &=& B\big(\cos{E} - e\big) + G\sqrt{1 - e^2}\sin{E}, 
\label{eq:xye}
\end{split}
\end{eqnarray}

\noindent
where $x_E$ and $y_E$ are the Cartesian coordinates of the projected orbit, $e$ is the planet's orbital
eccentricity, $E$ is the eccentric anomaly, and $A,\; B,\; F$ and $G$ are the modified Thiele-Innes 
constants given by

\begin{eqnarray}
\begin{split}
A &=& \frac{a}{d}\, (\cos{\omega} \cos{\Omega} - 
     \sin{\omega}\, \sin{\Omega}\, \cos{i}), \\ 
B &=& \frac{a}{d}\, (\cos{\omega} \sin{\Omega} + 
     \sin{\omega}\, \cos{\Omega}\, \cos{i}), \\ 
F &=&- \frac{a}{d}\, (\sin{\omega} \cos{\Omega} + 
     \cos{\omega}\, \sin{\Omega}\, \cos{i}) ,\\
G &=&- \frac{a}{d}\, (\sin{\omega} \sin{\Omega} - 
     \cos{\omega}\, \cos{\Omega}\, \cos{i}).
\end{split}
\end{eqnarray}

\noindent
In these equations, $d$ is the distance between the observer and the observed system in units of the planetary semimajor axis $a$.
We can rewrite equations (\ref{eq:xye}) in terms of the true anomaly $f$ as,

\begin{eqnarray}
\begin{split}
x_f &=& \frac{A}{a} \;r\; \cos{f} + \frac{F}{a}\; r\; \sin{f}, \\
y_f &=& \frac{B}{a} \;r\; \cos{f} + \frac{G}{a}\; r\; \sin{f}. 
\end{split}
\end{eqnarray}

\noindent
In these equations, \mbox{$r = a (1 - e^2)/(1 + e\cos{f})$} represents the planet's radial distance to its host star. 
Because the motion of the planet itself cannot be traced, 
we translate these equations into the apparent motion of the host star by the application of Newton's third law. That is,

\begin{eqnarray}
\begin{split}
x_\star &=& X - \mu x_f\,, \\
y_\star &=& Y - \mu y_f\,.
\end{split}
\end{eqnarray}

\noindent
Here, $X$ and $Y$ are the projected coordinates of the center of mass of the planet-star system, 
and $\mu$ denotes the planet-star mass-ratio as defined for equation (2).  

Assuming without the loss of generality that the barycenter of the star-planet system coincides with the 
origin of the associated coordinate system, the distance of the projected stellar orbit to 
the coordinate center will be equal to

\begin{eqnarray}
\begin{split}
{\rho^2} &= x_\star^2+y_\star^2 \\
&={\frac{\mu^2 a^2}{d^2}}\,\frac{{(1 - e^2)^2} \big[1 - \sin^2{i} \sin^2(f + \omega)\big]}{(1 + e \cos{f})^2}.\label{eq:rho2}
\end{split}
\end{eqnarray}

\noindent
The right-hand side of equation (\ref{eq:rho2}) is independent of $\Omega$ and has a global 
maximum at \mbox{$f=\pi ,\omega=0$} when $e=e^{max}$. This translates into a maximum astrometric amplitude given by

\begin{equation}
\rho^{max} = \rho^{circ}\> (1 + e^{max})\,,
\label{eq:rho}
\end{equation}

\noindent
where
\begin{equation}
\rho^{circ} = \frac{\mu a}{d}\,.
\end{equation}

\begin{figure}
\center
\includegraphics[angle=-90, scale=0.36]{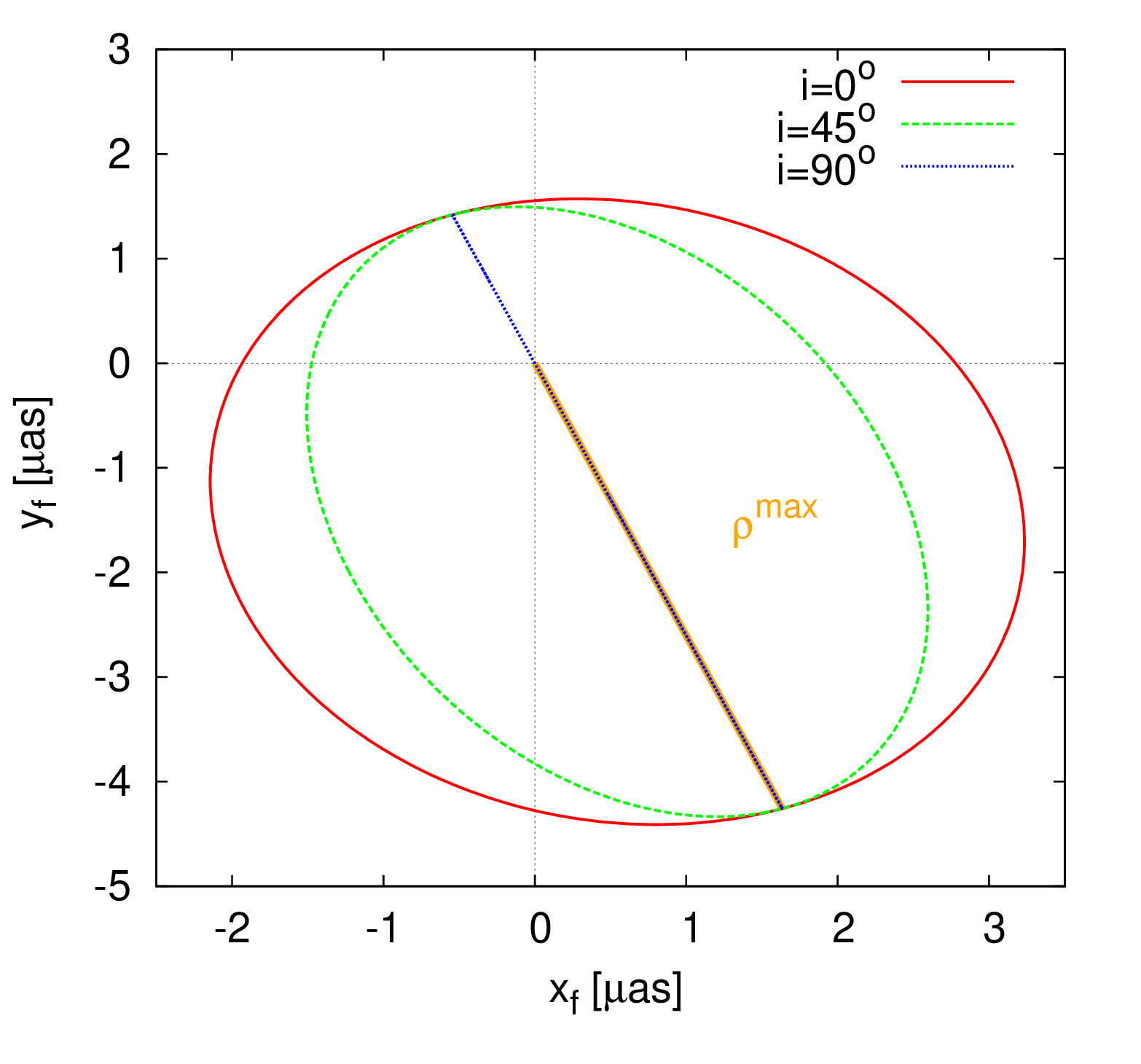}
\includegraphics[angle=-90, scale=0.36]{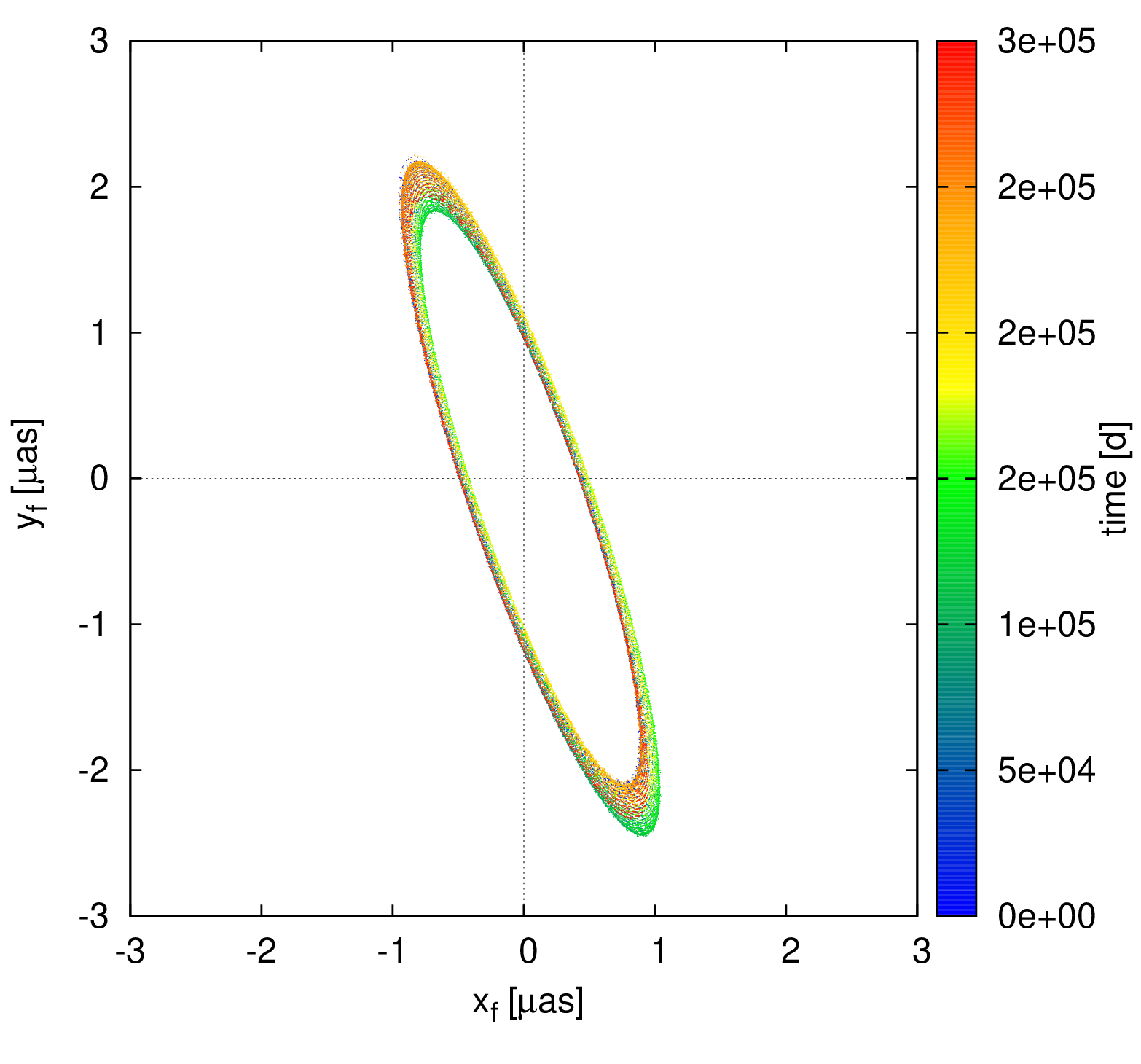}
\caption{\textit{Top:} The maximum astrometric amplitude, 
\mbox{$\rho^{max}=\mu\,{({x_f^2+y_f^2})^{1/2}}|_{f=\pi,\omega=0}$},
due to an Earth-like planet orbiting its Sun-like host star. The planet's orbital elements are  
$a=1$ AU, $e=e^{max}=0.5$, $\omega=0$, $\Omega=111^\circ$. As shown here, the maximum distance from the origin of the coordinate system 
is independent of the system's inclination with respect to the plane of the sky ($i$). \textit{Bottom:} Evolution of the astrometric signal ($x_f,y_f$) 
caused by an Earth-like planet in a binary star system. The planet is orbiting $\alpha$ Cen B at a distance of 1 AU. 
The evolution of the astrometric signal is shown for 3750 periods of $\alpha$ Centauri AB.
Since the system is coplanar, the changes in orientation and shape of the projected ellipse are due to variations in the planet's eccentricity ($e$) and argument of pericenter ($\omega$).  
\label{fig2}} 
\end{figure}

\noindent
The planetary maximum AM signal will again lie between $\rho^{circ}$ and $\rho^{max}$.
A remarkable feature of $\rho^{max}$ and $\rho^{circ}$ is their independence of the system's inclination $i$.
This is visualized in figure \ref{fig2}. The same figure also shows the time evolution of the AM signal due to an Earth-like
planet orbiting $\alpha$ Cen B at a distance of 1 AU. 

The astrometric RMS values are given by   

\begin{eqnarray}
\begin{split}
\langle\langle  \rho  \rangle\rangle_M =& \rho^{circ} \left[1+\frac{3 \langle e^2 \rangle_M}{2} \;+ \right.\\
&\left. \left(-\frac{1}{2}+\frac{\langle e^2\rangle_M}{4} (5 \cos[2 \omega ]-3)\right) \sin^2{i}\right]^{1/2}, \label{eq:rhorms}
\end{split}
\end{eqnarray}

\noindent
and

\begin{eqnarray}
\begin{split}
 \langle\langle \rho \rangle\rangle_{M,\omega} =& \frac{\rho^{circ}}{2}\left[3 + \frac{9}{2} \langle e^2 \rangle_{M,\omega} \; + \right.\\
         &\left. \left(1 + \frac{3}{2} \langle e^2 \rangle_{M,\omega} \right) \cos(2 i)\right]^{1/2}.\label{eq:rhorms2}
\end{split}
\end{eqnarray}

\noindent
Details regarding the derivation of equations (\ref{eq:rhorms}) and (\ref{eq:rhorms2}) can be found in appendix \ref{sec:ave2}.
In contrast to the doubly averaged equations for the RMS of an RV signal, equation (\ref{eq:rhorms2}) 
shows a dependence on the binary's eccentricity. In cases where the planetary inclination $i$ coincides 
with the inclination of the binary itself, analytic expressions  
for $\langle e^2 \rangle_{M,\omega}$ are available \citep{georgakarakos-2003,georgakarakos-2005}\footnote{The analytic expressions given 
in these articles are also averaged over initial phases, i.e. different relative starting positions of the planet and the binary stars.}. 

\section{Transit Photometry} \label{sec:tp}
In transit photometry, signal strength is equivalent to the relative depth of the dint the planet produces in the stellar light-curve during its transit.
Assuming that the star-planet configuration allows for occultations, 
and excluding grazing transits,
the fractional depth of the photometric transit ($TD$) produced by an Earth-like 
planet is simply given by the proportion of the luminous area of the disk of the star that is covered
by the planet as the planet moves between the observer and the star. Ignoring limb darkening, 
that means, $TD\simeq R_p^2 / R_\star^2$ 
where $R_p$ is the radius of the planet and $R_\star$ is the stellar radius.
The overall probability to observe a transit is given by \citep{borucki-summers-1984}:

\begin{equation} 
p_T=\frac{R_\star}{r_T} \,.
\label{eq:prob}
\end{equation}

\noindent
In equation (\ref{eq:prob}), $r_T$ is the radial distance of the planet to the star during the transit. 
For an eccentric 
planetary motion, the planet-star distance during transit can be expressed as 
\mbox{$r_T\simeq a(1-e^2)/(1+e \cos{\bar{\omega}})$}  \citep{ford-et-al-2008}, where $\bar{\omega}$ denotes
the argument of pericenter measured from the line of sight\footnote{Note that this is different from 
the conventions used for RV and AM measurements.}. 
In analogy to sections \ref{sec:rv} and \ref{sec:astrometry}, the maximum and averaged transit
probability for a planet perturbed by the secondary star in a planar configuration can be calculated by
substituting for $r_T$ in equation (\ref{eq:prob}) and averaging over $\bar{\omega}$.
This will result in 

\begin{equation} 
 p_T^{max} \simeq \frac{R_\star}{a(1-e^{max})}\,,
\label{eq:prob3}
\end{equation}

\noindent
and

\begin{equation}
\langle p_T \rangle_{\bar{\omega}} \simeq  \frac{R_\star}{a(1-\langle e^2\rangle_{M,\omega})}\,.
\label{eq:prob4}
\end{equation}

\noindent
Equations (\ref{eq:prob3}) and (\ref{eq:prob4}) indicate that the increase in the eccentricity of the  planet due to the 
perturbation of the secondary increases the probability of transit.
In deriving these equations, we have ignored the occultation of the planet by the
second star. However, depending on the period ratio between the secondary and the planet,
such conjunctions are either scarce or short-lived. 
Consequently, their contribution to the probability of witnessing a planetary transit is negligibly small.

\section{Application to the $\alpha$ Centauri system}\label{sec:results} 

In this section we will show that the previously derived analytic expressions produce results that are in good agreement 
with the current observations of $\alpha$ Cen Bb. We will also present numerical evidence that the presence - or absence - of
an additional terrestrial-planet in the HZ of $\alpha$ Cen B cannot be derived easily from the orbit evolution of $\alpha$ Cen Bb.
%since the mutual interactions between such planets is small compared to the gravitational influence of the binary.
Consequently, we argue that an independent detection of additional terrestrial companions might be difficult, but more promising.  
For this purpose, we will determine the HZ of $\alpha$ Cen B, as well as the RV, AM and 
TP signatures of an Earth-like planet orbiting in the HZ of $\alpha$ Cen B. 
Since there is no a priori reason why the brighter component of $\alpha$ Centauri could not be hosting a terrestrial planet as well, we perform a similar
analysis for $\alpha$ Cen A.
We will also study the behavior of equations (\ref{eq:vr}-\ref{eq:prob4}) for a broad range of binary 
eccentricities.

\subsection{$\alpha$ Centauri's terrestrial planet}
The planet discovered around $\alpha$ Cen B offers a perfect opportunity to compare the RV amplitude predictions derived in 
section \ref{sec:rv} with actual measurements. The planet's known orbital parameters are given in Table~\ref{tab1}.
In Table~\ref{tab2} we present the analytic estimates of section \ref{sec:rv} applied to the \mbox{$\alpha$ Centauri ABb} system. 
Assuming the system to be coplanar ($i\simeq 79.2^\circ$), the predicted RV amplitude for circular planetary motion ($V_r^{circ}$) is very close to the observed RV amplitude.
This is not surprising, since the planetary parameters were derived from an RV signal using the same methodology in reverse. 
While still well within measurement uncertainties, the deviation of the maximum RV amplitude ($V_r^{max}$) from the observed value 
is larger than that of $V_r^{circ}$. On the one hand, this might indicate that the 
planet is currently in an orbital evolution phase where its eccentricity is almost zero.
On the other hand, the planet may be too close to its host star for our model to predict $V_r^{max}$ correctly.
In fact, we show in section \ref{sec:cenhz} that the latter explanation is more likely, 
since the influence of general relativity (GR) cannot be neglected in this case.
Estimates based on Newtonian physics exaggerate the actual eccentricity of $\alpha$ Cen Bb. Its orbit 
remains practically circular despite the interaction with the binary star (see section \ref{sec:cenhz} for a detailed discussion).
This justifies the assumption of a circular planetary orbit made by \citet{dumusque2012}.

Since we are especially interested in additional habitable planets, however, 
it is worthwhile to ask whether predictions on the orbital evolution of $\alpha$ Cen Bb 
can be used to exclude the presence of other gravitationally active bodies in the system. 
In other words: Could an Earth-like planet still orbit in the HZ of $\alpha$ Cen B or would the accompanying distortions of the orbit of $\alpha$ Cen Bb be significant 
enough to detect them immediately?
Before we try to answer these questions, we need to briefly recall some important aspects regarding HZs in binary star systems. 

\subsection{Classification of HZs}\label{sec:hz}
Combining the classical definition of a HZ \citep{kasting-et-al-1993} with the dynamical properties of 
a planet-hosting double star system, \citet{eggl-et-al-2012} have shown that one can
distinguish three types of HZ in an S-type binary system:

\begin{description}

\item[The Permanently Habitable Zone (\textbf{PHZ})]
where a planet \textit{always} stays within the insolation limits ($S_I$, $S_O$) as defined by 
\citet{kasting-et-al-1993} and \citet{underwood-et-al-2003}. In other words, despite the changes in 
its orbit, the planet never leaves the classical HZ. The total insolation the planet 
receives will vary between the inner $(S_I)$ and outer $(S_O)$ effective radiation limits as 
$S_I \geq S_{tot} \geq S_O$ where, for a given stellar spectral type, 
$S_I$ and $S_O$ are in units of Solar constant (1360 [W/m$^2$]).

\item[The Extended Habitable Zone (\textbf{EHZ})] where,
in contrast to the PHZ, parts of the planetary orbit may lie outside the HZ 
due to the planet's high eccentricity, for instance. Yet, the binary-planet configuration is still 
considered to be habitable when most of the planet's orbit remains inside the boundaries of the HZ. 
In this case, $\left \langle S_{tot}\right \rangle_t +\sigma \leq S_I \quad$ and 
$\quad \left \langle S_{tot}\right \rangle_t -\sigma \geq S_O \label{eq:ehzlim}$
where  $\left \langle S_{tot} \right \rangle_t$ denotes the time-averaged effective insolation from 
both stars and $\sigma^2$ is the effective insolation variance.   

\item[The Averaged Habitable Zone (\textbf{AHZ}).]
Following the argument of \citet{williams-pollard-2002} that planetary eccentricities up to $e<0.7$ 
may not be prohibitive for habitability as long as the atmosphere can act as a buffer, the AHZ is 
defined as encompassing all configurations which support the planet's \textit{time-averaged} 
effective insolation to be within the limits of the classical HZ. Therefore,
$S_I \geq  \left \langle S_{tot} \right \rangle_t \geq S_O$.

\end{description}

\noindent
 Analytic expressions for the maximum insolation, the average insolation $(\langle S_{tot} \rangle)$, 
and insolation variance that a planet encounters in a  binary system have been derived in \citet{eggl-et-al-2012}. 
We refer the reader to that article for more details. 

\begin{figure}
\begin{tabular}{cc} $\alpha$ Cen A & $\alpha$ Cen B \\
\includegraphics[angle=-90, scale=0.36]{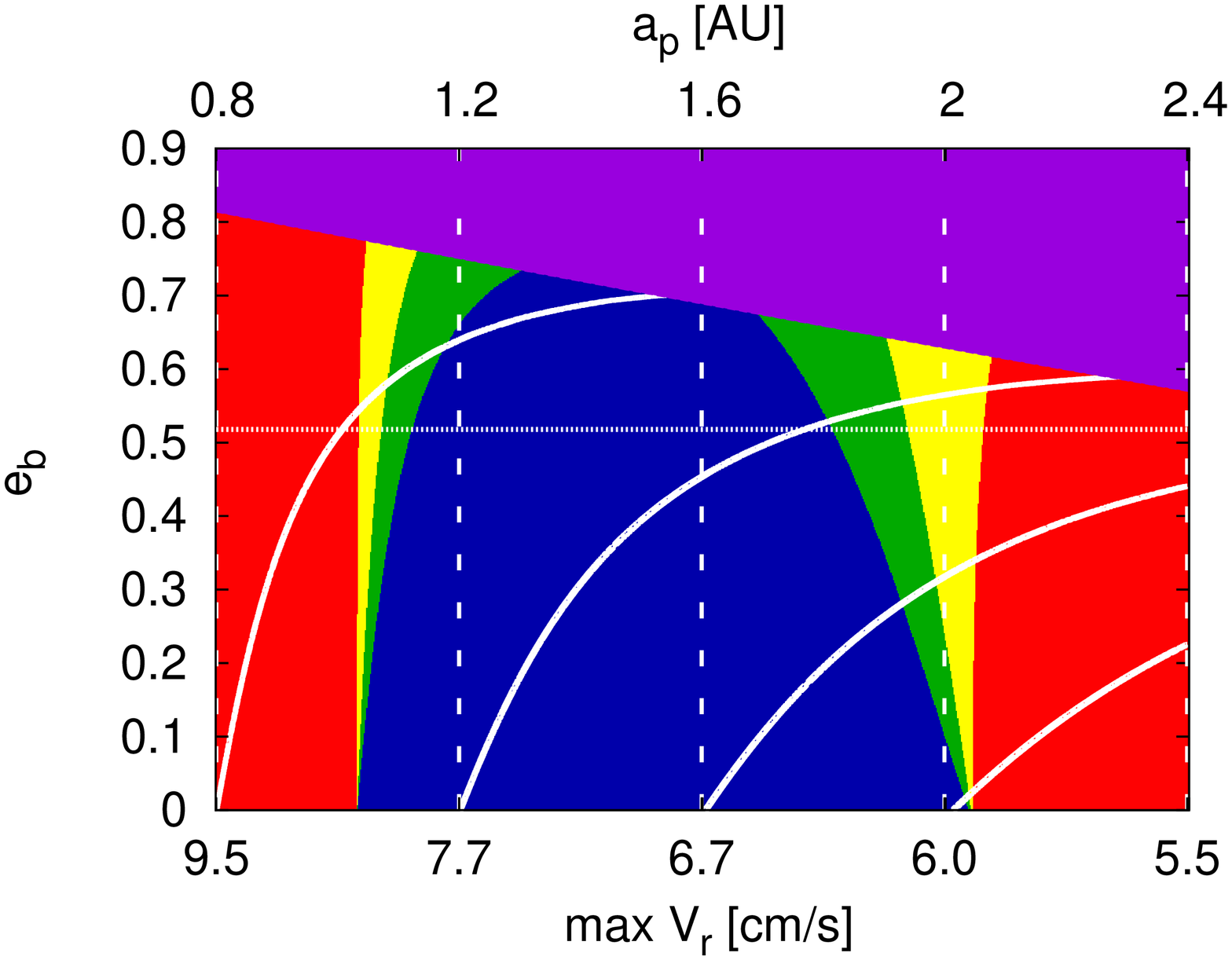} &
 \includegraphics[angle=-90, scale=0.36]{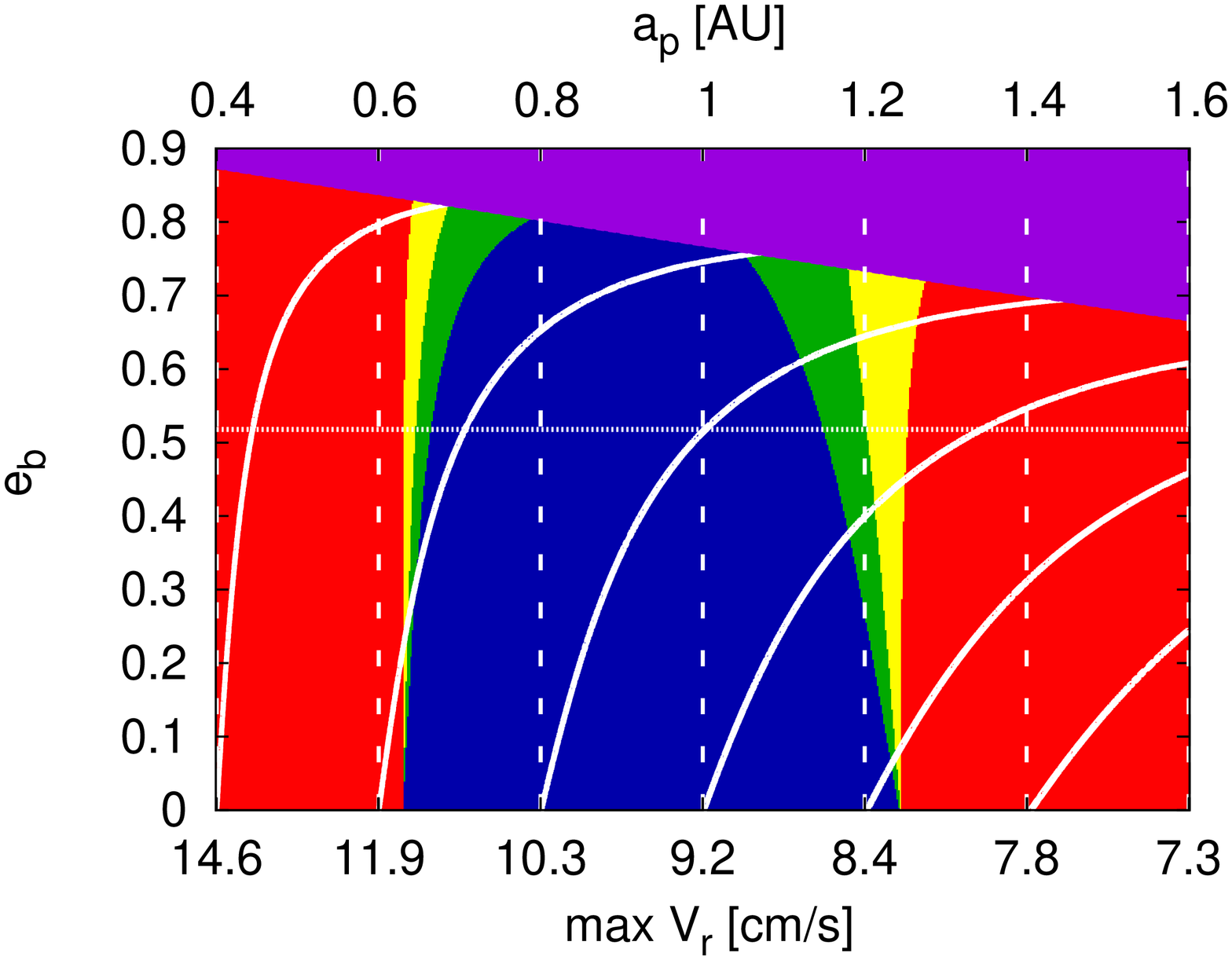}\\
 \includegraphics[angle=-90, scale=0.36]{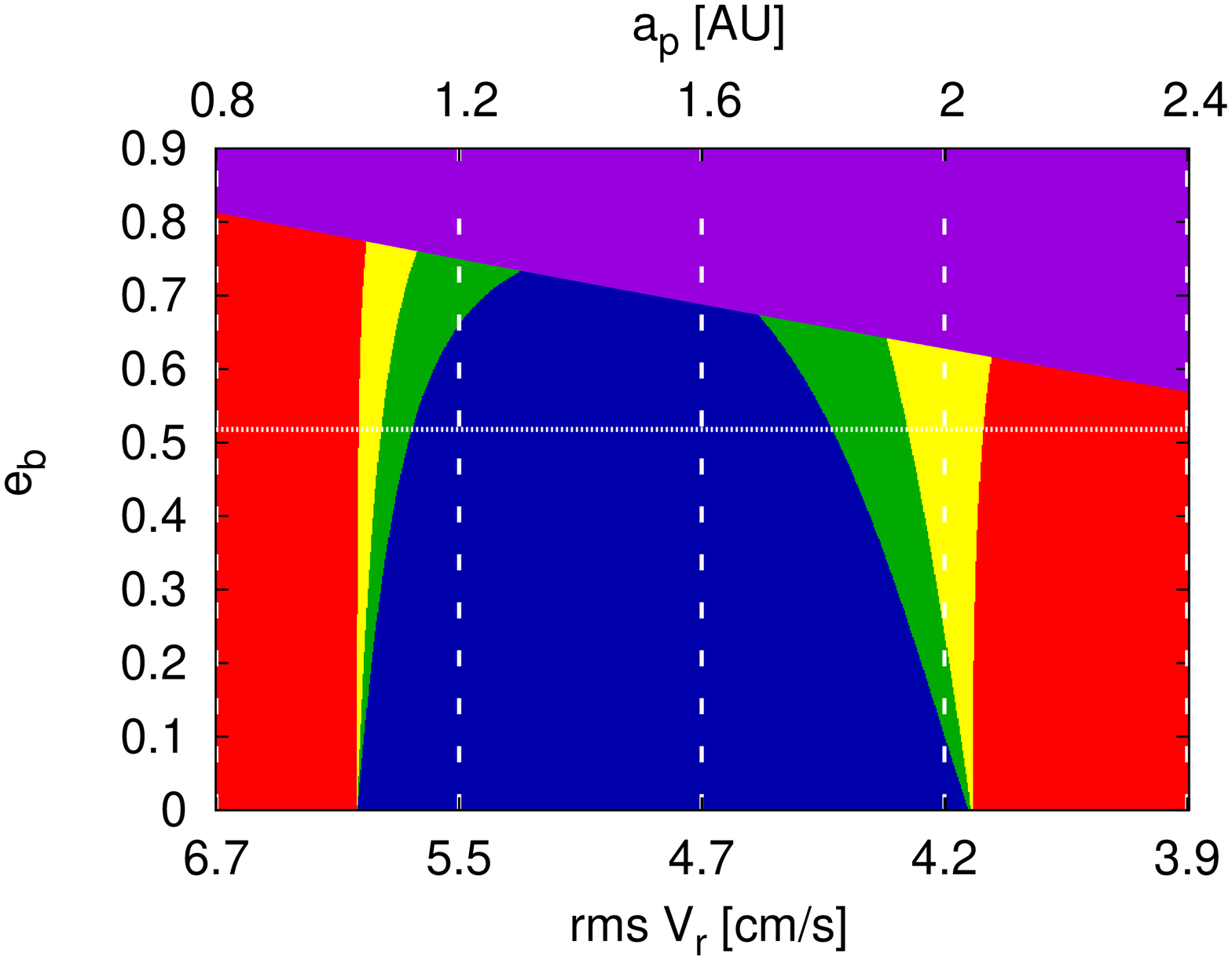}&
\includegraphics[angle=-90, scale=0.36]{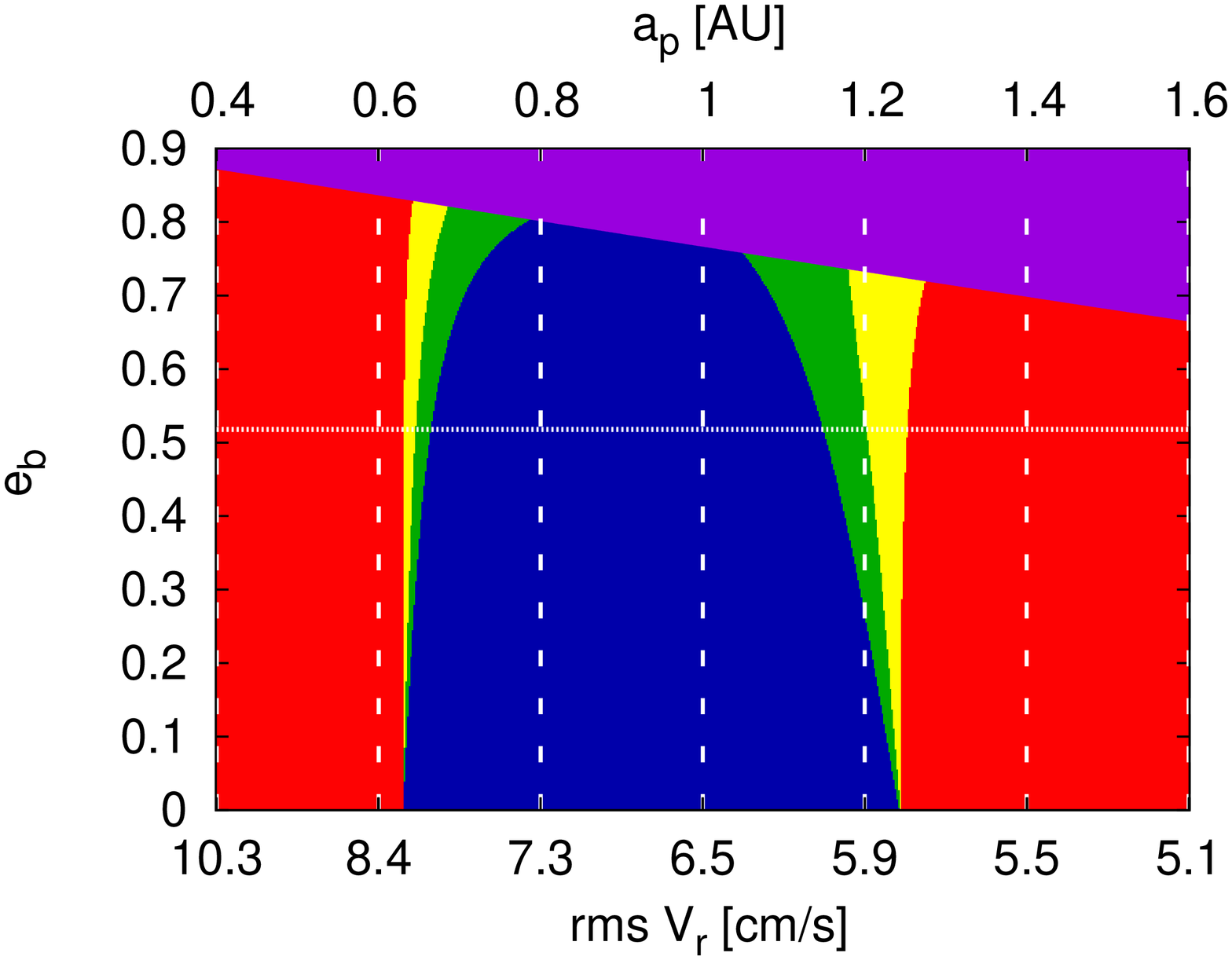}\\
\end{tabular}
\caption{Habitability maps showing the  maximum amplitudes and RMS of the RV signal of the planet-hosting stars 
for the $\alpha$ Centauri system. The quantity $a_p$ is the semimajor axis of the terrestrial planet
and $e_b$ is the binary eccentricity. The color blue shows the PHZ, the EHZ is green, yellow indicates 
the AHZ, and red means that the planet is outside of any defined HZ. The purple area denotes 
dynamical instability.  The horizontal line in each panel denotes the actual eccentricity 
of the $\alpha$ Centauri binary. As shown here, $V_r^{max}$ reacts strongly to enhanced binary eccentricities 
(\textit{top row}, curved, solid lines) whereas in contrast, $\langle\langle V_r\rangle\rangle_{M,\omega}$ is independent 
of the binary's eccentricity (\textit{bottom row}, straight, vertical lines). The straight, vertical lines in the top row correspond to RV amplitudes for circular orbits ($V_r^{circ}$). 
% Planetary orbits with 
% dynamically elevated eccentricities allow for larger semimajor axes compared to circular orbits (dashed, vertical lines) while still producing
% similar RV peak amplitudes. This can be seen from the solid, white lines that
% represent $V_r^{max}$ iso-curves. 
See section \ref{sec:detect} for details.
\label{fig3}}
\end{figure}

\begin{figure}
\begin{tabular}{cc} $\alpha$ Cen A & $\alpha$ Cen B\\
\includegraphics[angle=-90, scale=0.36]{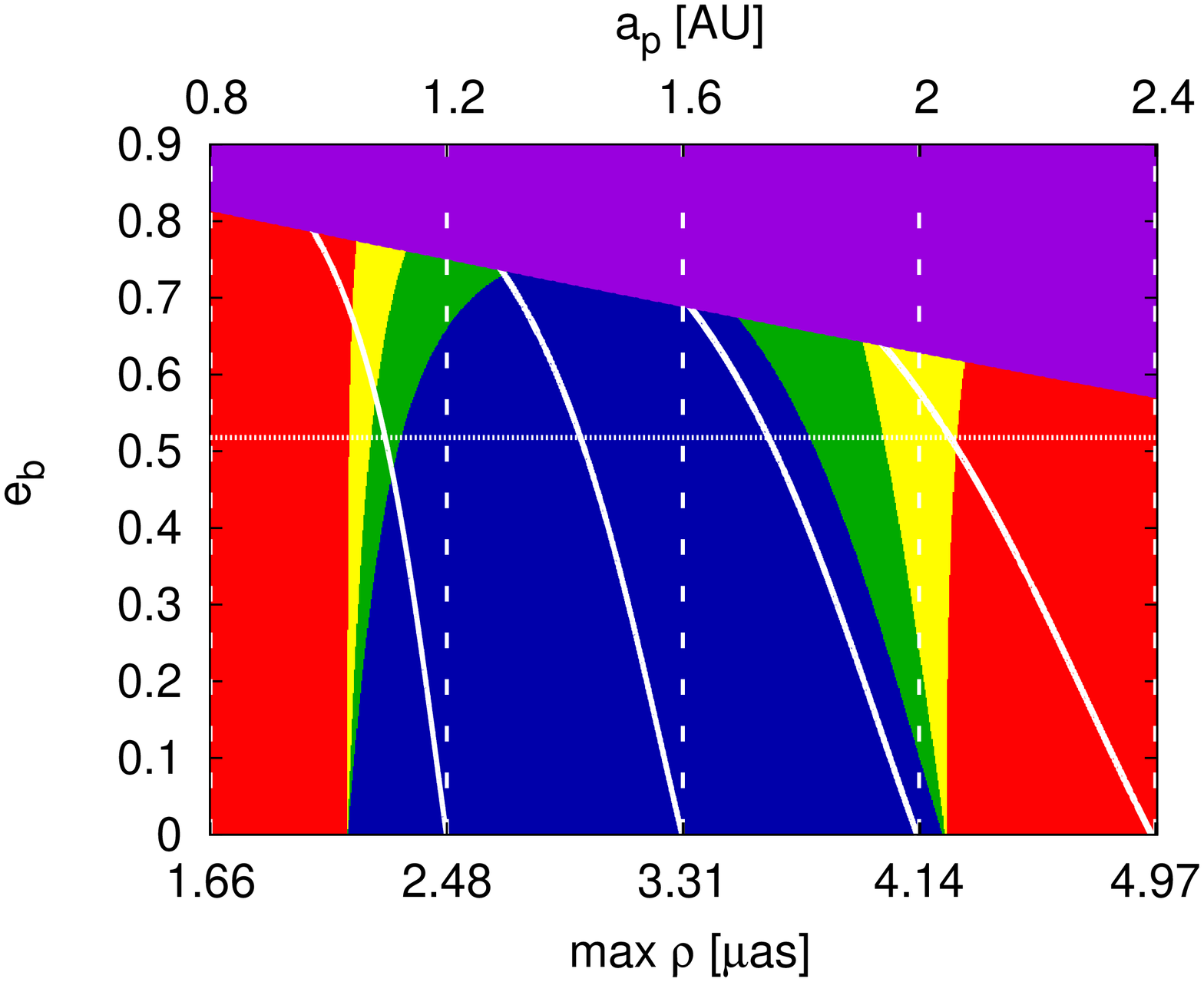} &
 \includegraphics[angle=-90, scale=0.36]{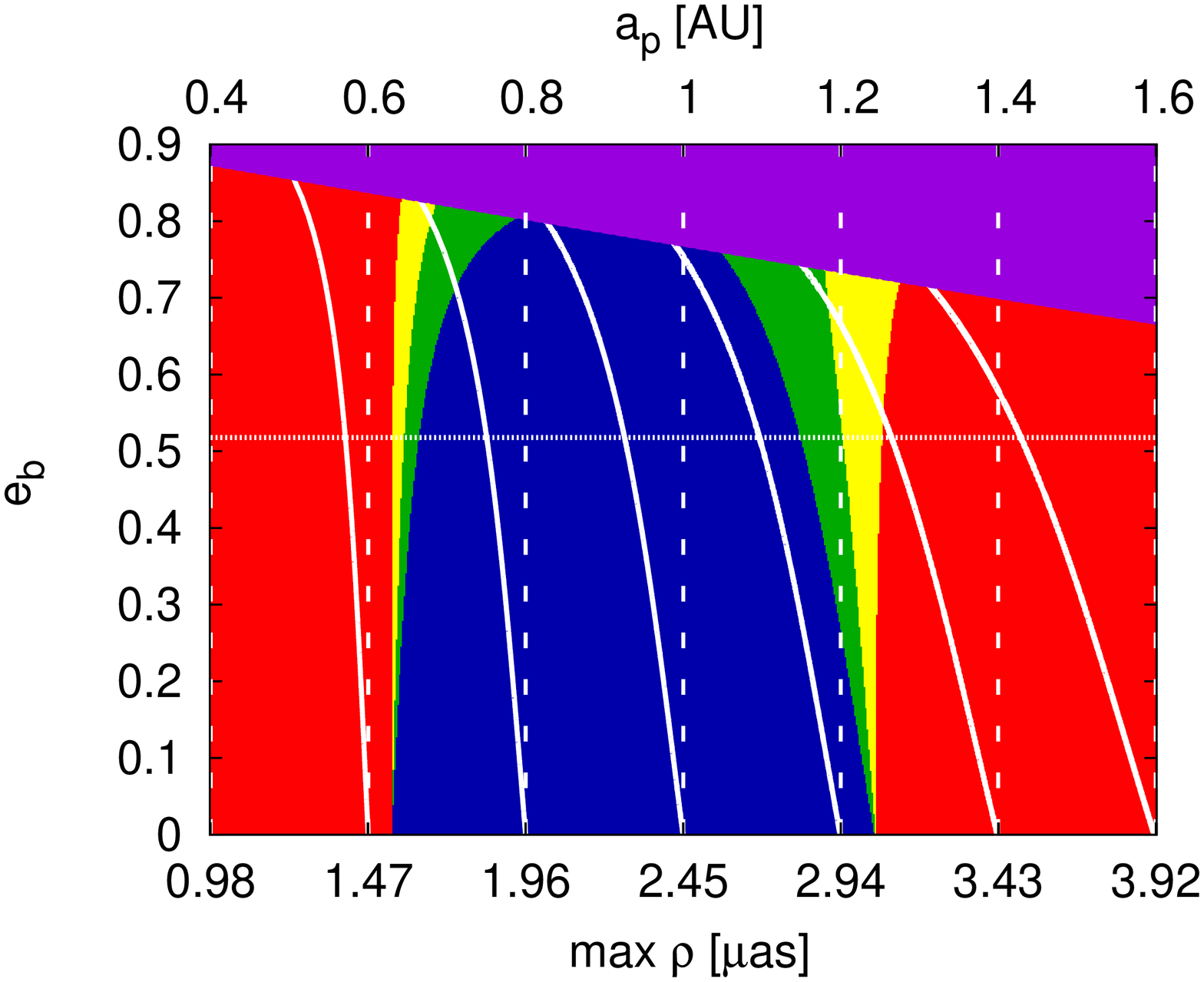}\\
 \includegraphics[angle=-90, scale=0.36]{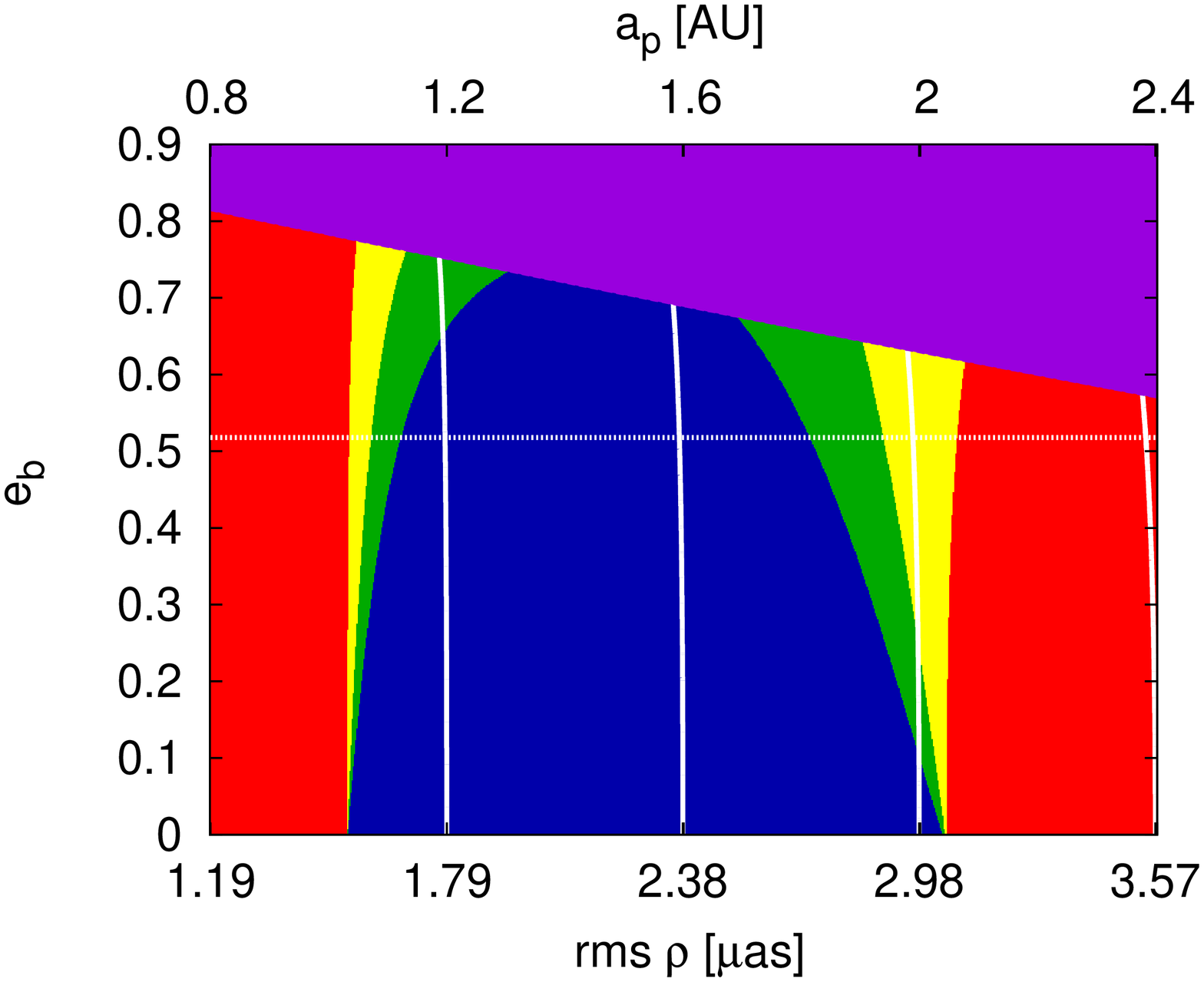}&
\includegraphics[angle=-90, scale=0.36]{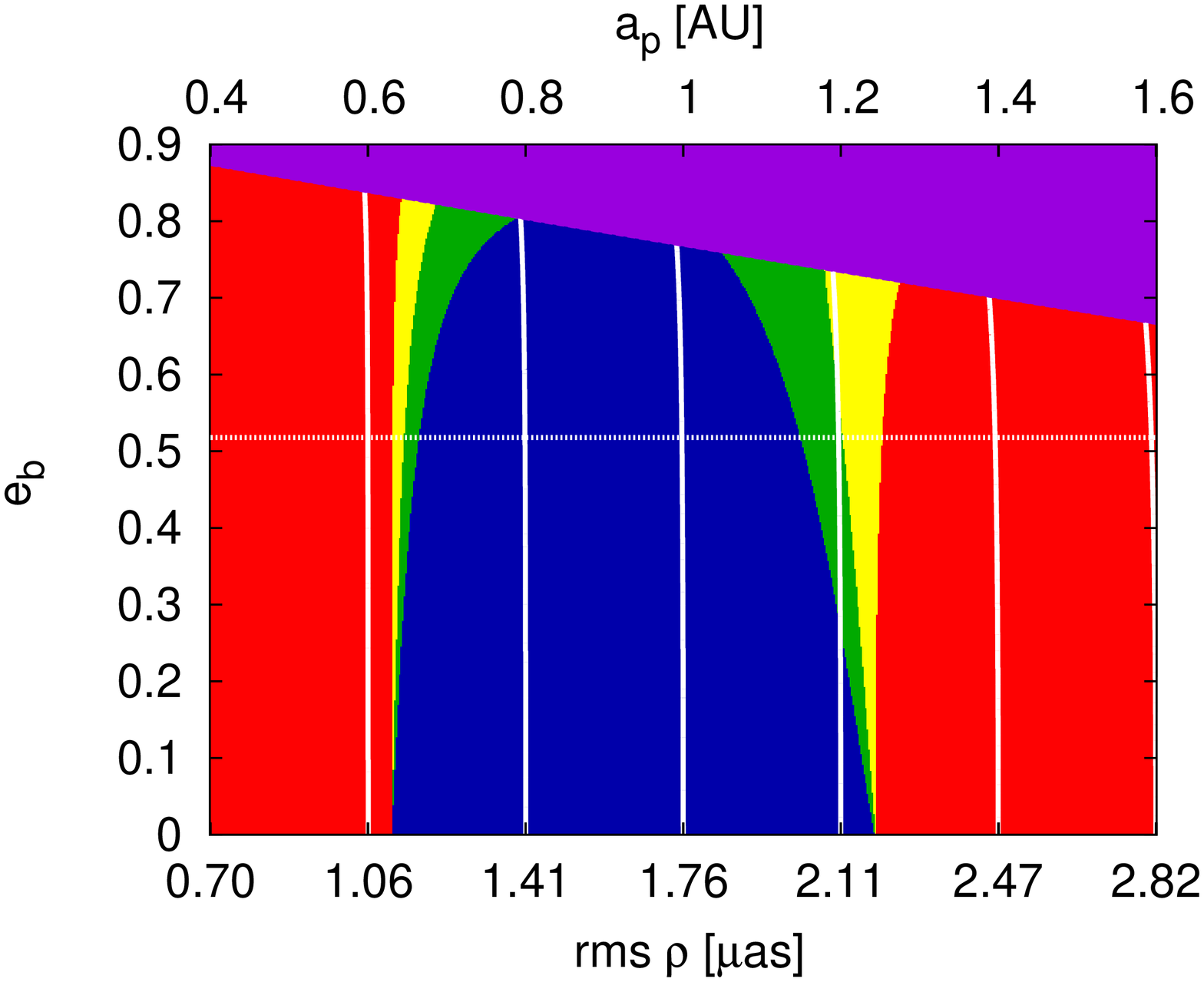}\\
\end{tabular}
\caption{Habitability maps showing the  maximum amplitudes and RMS of the astrometric signals
for the $\alpha$ Centauri system. The color coding is similar to figure \ref{fig3}.
The vertical dashed lines in the top panels represent regions with 
similar values of $\rho^{circ}$. The curved lines in these panels show regions with similar $\rho^{max}$ 
amplitudes. In the bottom panels, the vertical lines represent areas of equal RMS amplitudes, 
$\langle\langle \rho \rangle\rangle_{M,\omega}$. One can see that planetary orbits with 
dynamically enhanced eccentricities can have smaller semimajor axes and still produce similarly 
high astrometric amplitudes as circular orbits which are more distant from the host-star.  
\label{fig4}} 
\end{figure}

Figures \ref{fig3} and \ref{fig4} show the application of the proposed habitability classification 
scheme to the $\alpha$ Centauri system. In these figures, blue denotes PHZs, green shows EHZs, and yellow 
corresponds to AHZs. The red areas in figures \ref{fig3} and \ref{fig4} are uninhabitable, and purple stands for dynamically 
unstable regions. Table \ref{tab1} shows the physical properties of the system.
We used the formulae by \citet{underwood-et-al-2003} 
%and \citet{Selsis-et-al-2007} 
to calculate $S_{I}$ and $S_{O}$ for the given effective temperatures of $\alpha$
Cen A and B. In general, these formulae allow for extending
the analytic estimates for HZs, as given by \citet{eggl-et-al-2012}, to main sequence stars 
with different spectral types. Runaway greenhouse and maximum greenhouse insolation limits were used to 
determine the inner and outer boundaries of HZ, respectively.

\begin{figure}
\begin{tabular}{cc} $\alpha$ Cen A & $\alpha$ Cen B \\
\includegraphics[angle=-90, scale=0.36]{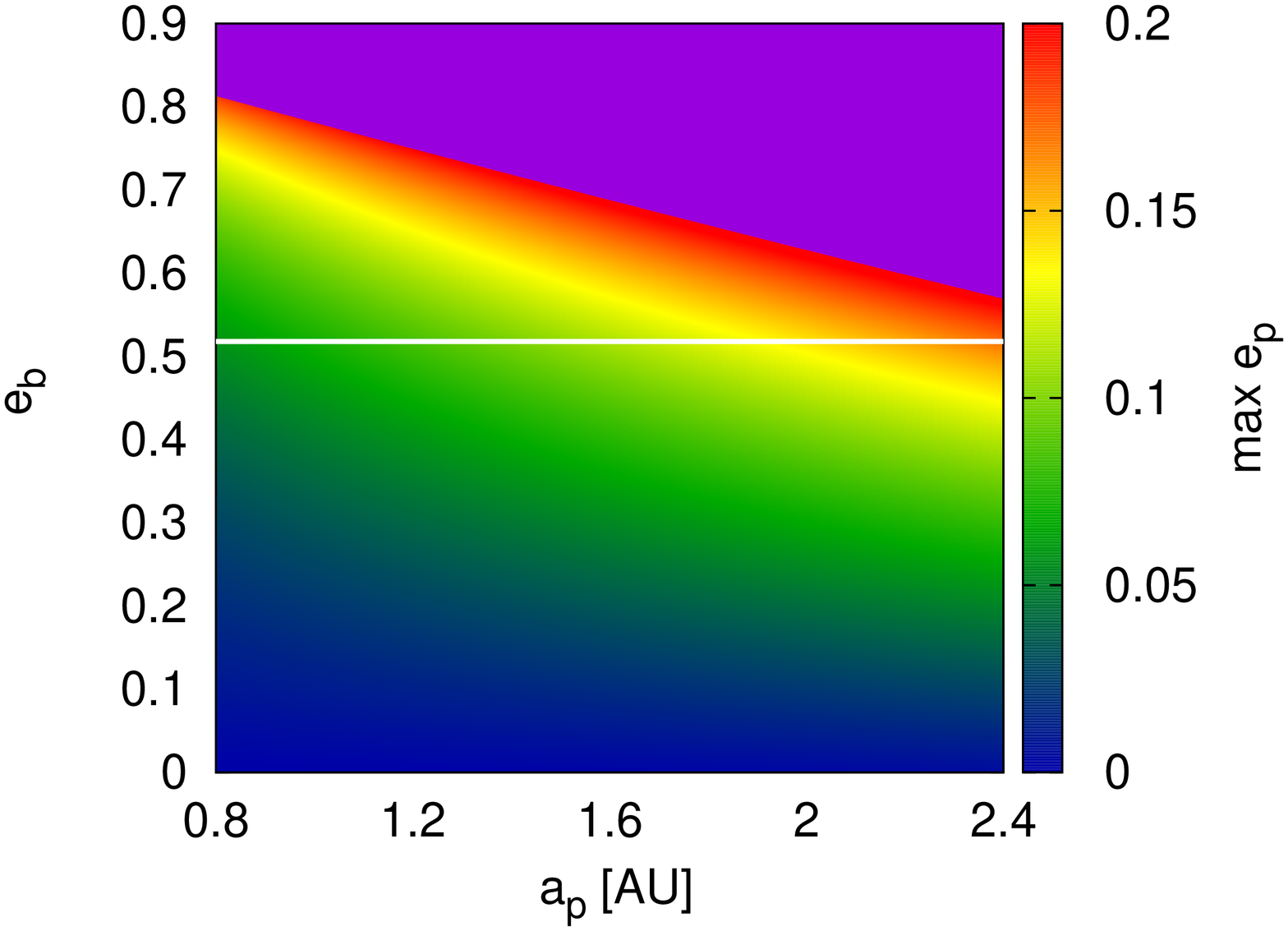} &
 \includegraphics[angle=-90, scale=0.36]{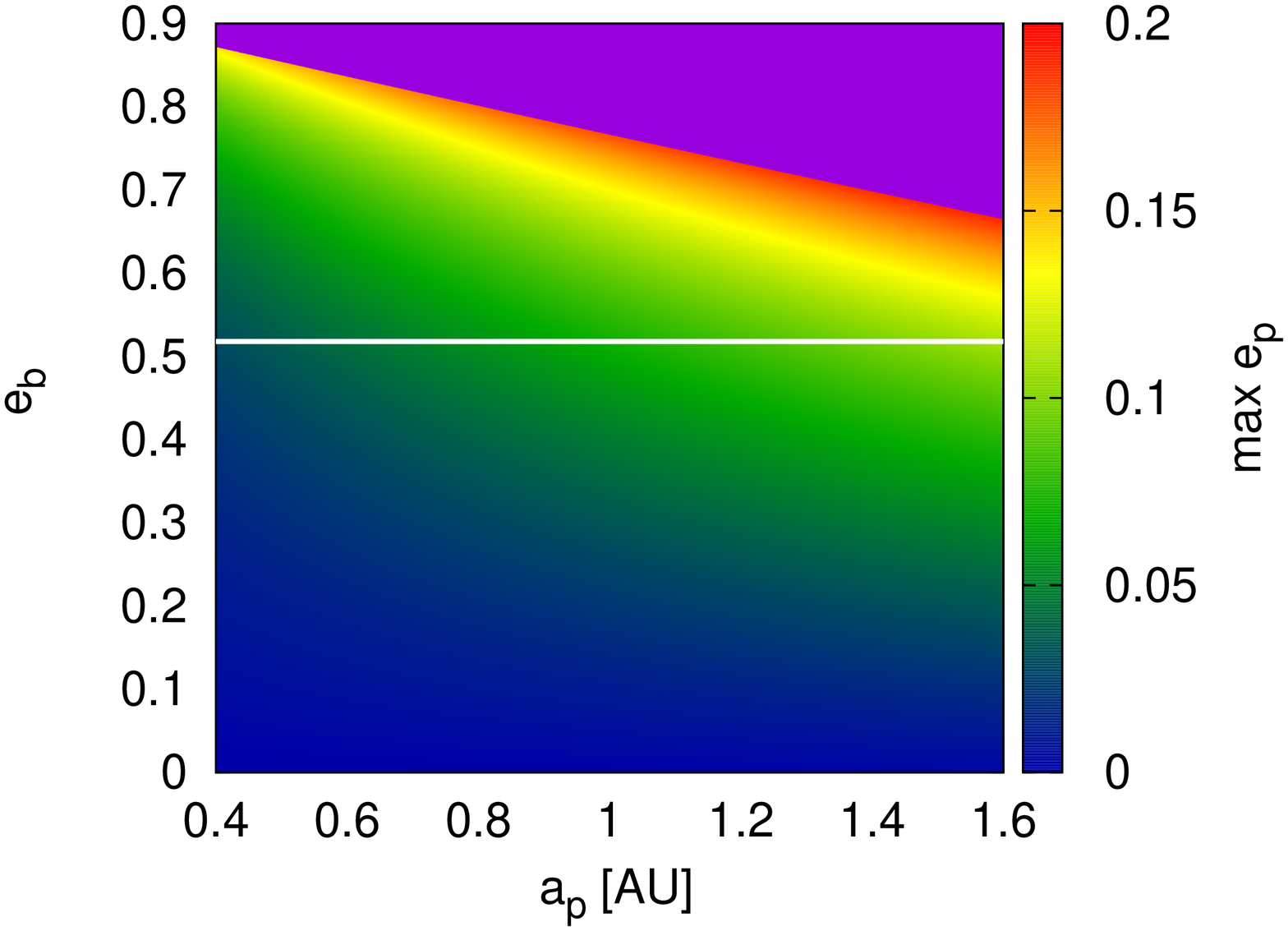}\\
 \includegraphics[angle=-90, scale=0.36]{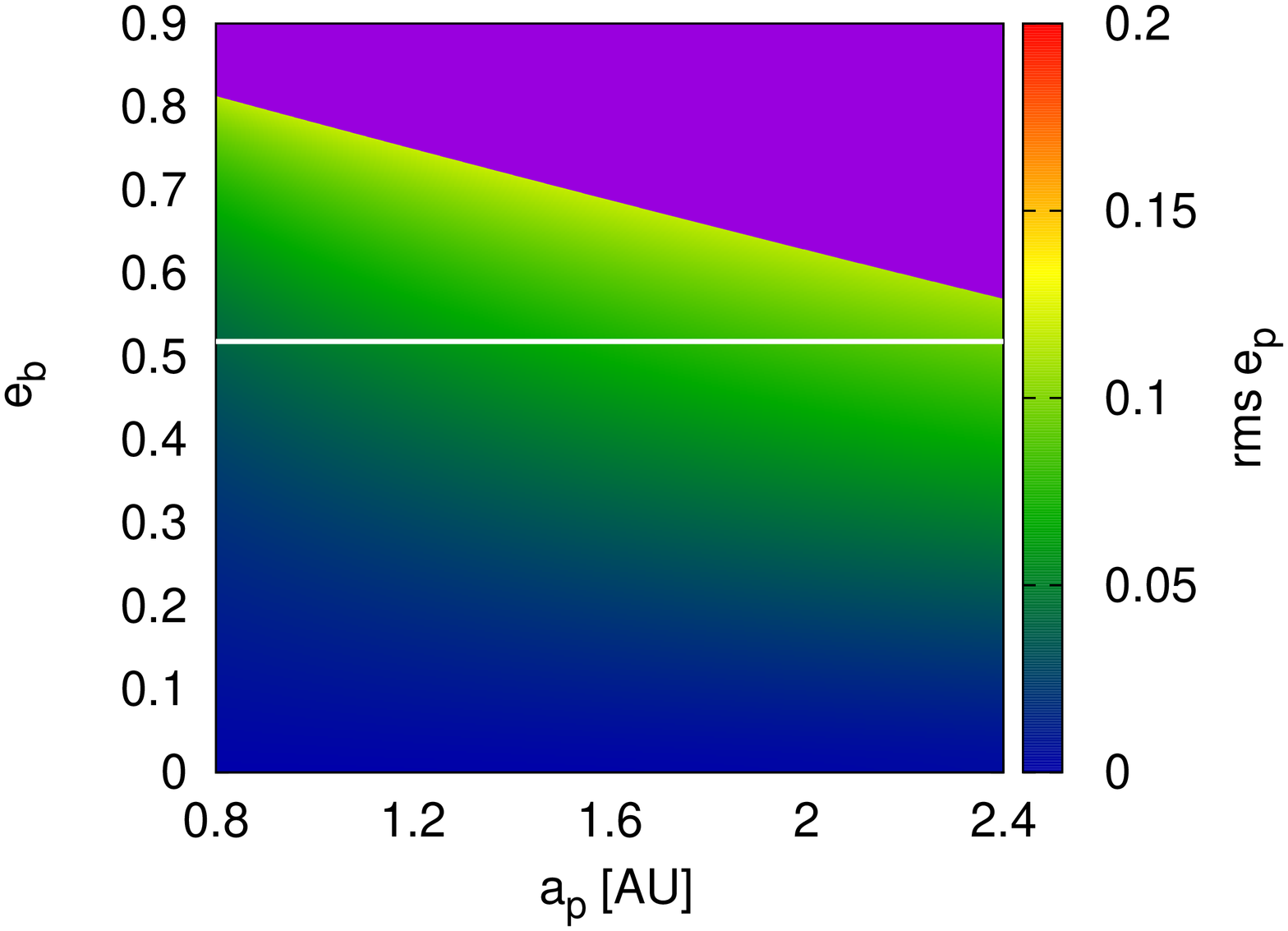}&
\includegraphics[angle=-90, scale=0.36]{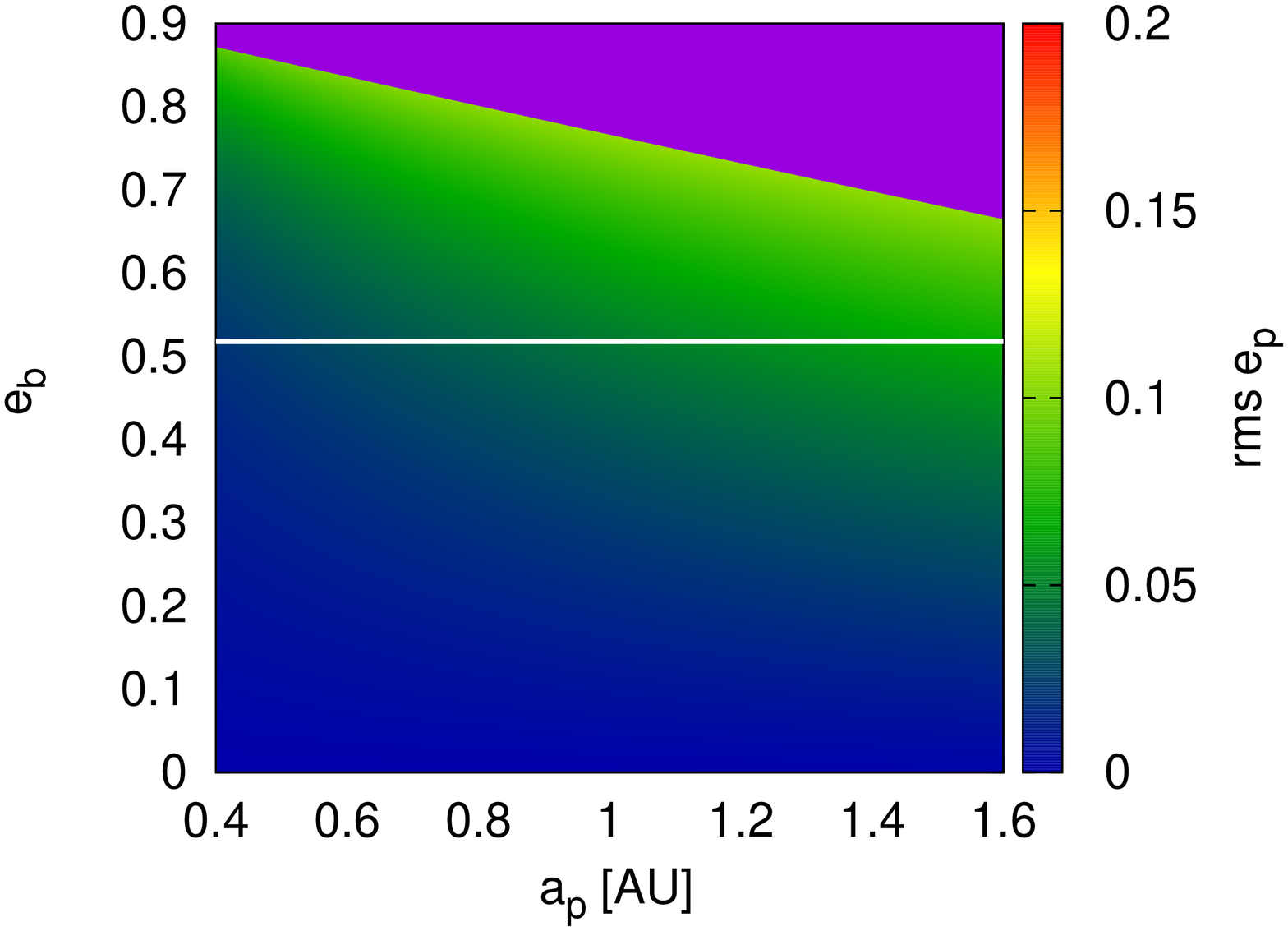}\\
\end{tabular}
\caption{Graphs of the maximum ($e_p^{max}$) and RMS ($\langle e_p^2\rangle_{M,\omega}^{1/2}$) values of the planetary eccentricity for different values of the
planet semimajor axis ($a_p$) and the binary eccentricity ($e_b$) for $\alpha$ Cen A (\textit{left}) and $\alpha$ Cen B 
(\textit{right}). The purple region denotes orbital instability. The horizontal line indicates the actual eccentricity of 
the $\alpha$ Centauri binary. 
\label{fig5}} 
\end{figure}

As shown in figures \ref{fig3} and \ref{fig4}, the locations of the HZs and the detectability of habitable planets in those regions
depend strongly on the eccentricity of the binary ($e_b$). The actual eccentricity of the $\alpha$ Centauri 
system is denoted by a horizontal line at $e_b=0.5179$. The values for the borders 
of the different HZs using $\alpha$ Centauri's actual eccentricity are listed in Table \ref{tab3}. 
As shown here, both stars permit dynamical stability for habitable, Earth-like planets. 
Due to the difference in stellar luminosities, the HZs 
around $\alpha$ Cen A are larger and farther away from the host star compared to  $\alpha$ Cen B. 
Since the binary's mass-ratio is close to 0.45, the gravitational 
influence of  $\alpha$ Cen B is more pronounced on the PHZ of $\alpha$ Cen A. 
This is a consequence of the larger injected planetary eccentricities ($e_p$) as can be seen from the top row of figure \ref{fig5}.
The relatively larger gravitational influence of $\alpha$ Cen B onto 
the HZ of $\alpha$ Cen A is also mirrored in the fact that the region of dynamical instability (purple) 
reaches towards lower binary eccentricities. 
The change in the range and configuration of HZs with the change in planetary semimajor axis and eccentricity
of the binary is pronounced.
A clear shrinking trend for PHZ and EHZ can be observed for high values of the binary's eccentricity. While as shown by \citet{eggl-et-al-2012}, the AHZ in general expands slightly when the eccentricity
is enhanced, figures \ref{fig3} and \ref{fig4} show that in the $\alpha$ Centauri system, this HZ depends only weakly 
on $e_b$ making it the closest approximation to the classical HZ as defined by \citet{kasting-et-al-1993}. 
Comparing these results with the existing studies on the HZs for $\alpha$ Cen B such as \citet{guedes-et-al-2008} 
and \citet{forgan-2012}, one can see that the values of the inner boundaries of the HZs around 
$\alpha$ Cen B as given in figures \ref{fig3} and \ref{fig4}, coincide well with the previous studies. \citet{forgan-2012} 
even found a similar shrinking trend with higher planetary eccentricity. Yet, \citet{forgan-2012} did not take the actual coupling between
the planet's eccentricity and the binary's orbit into account. The limits for the outer boundaries of HZ in our model are different
from the ones in \citet{forgan-2012} since different climatic assumptions were made. 
In this work we used insolation limits for atmospheric collapse assuming a maximum greenhouse atmosphere 
\citep{kasting-et-al-1993}  whereas  \citet{forgan-2012} focused on emergence from snowball states.

\subsection{Additional terrestrial planets in $\alpha$ Centauri's HZs}\label{sec:cenhz}
While the classification of habitable zones presented in the previous section is globally applicable to binary star systems, 
the analytic estimates to calculate their extent \citep{eggl-et-al-2012} are only strictly valid for three body systems, 
e.g. the binary star and a planet. Additional perturbers will influence the shape and size of the HZs.
It is thus necessary to investigate which effect the already discovered planet around $\alpha$ Cen B would have on an additional
terrestrial planet in $\alpha$ Cen B's HZ.

If the mutual perturbations were large, the HZ boundaries given in Table~\ref{tab3}  would have to be adapted, but
 $\alpha$ Cen Bb's orbital evolution could also contain clues on the presence - or absence - of an additional planet.
Should the interaction between the inner planet and an additional terrestrial body in the HZ be small, then the 
HZ boundaries would hold. However, a detection of the habitable planet via its influence on $\alpha$ Cen Bb's orbit would become difficult.

\begin{figure}
\begin{tabular}{cc} 
\includegraphics[angle=-90, scale=0.34]{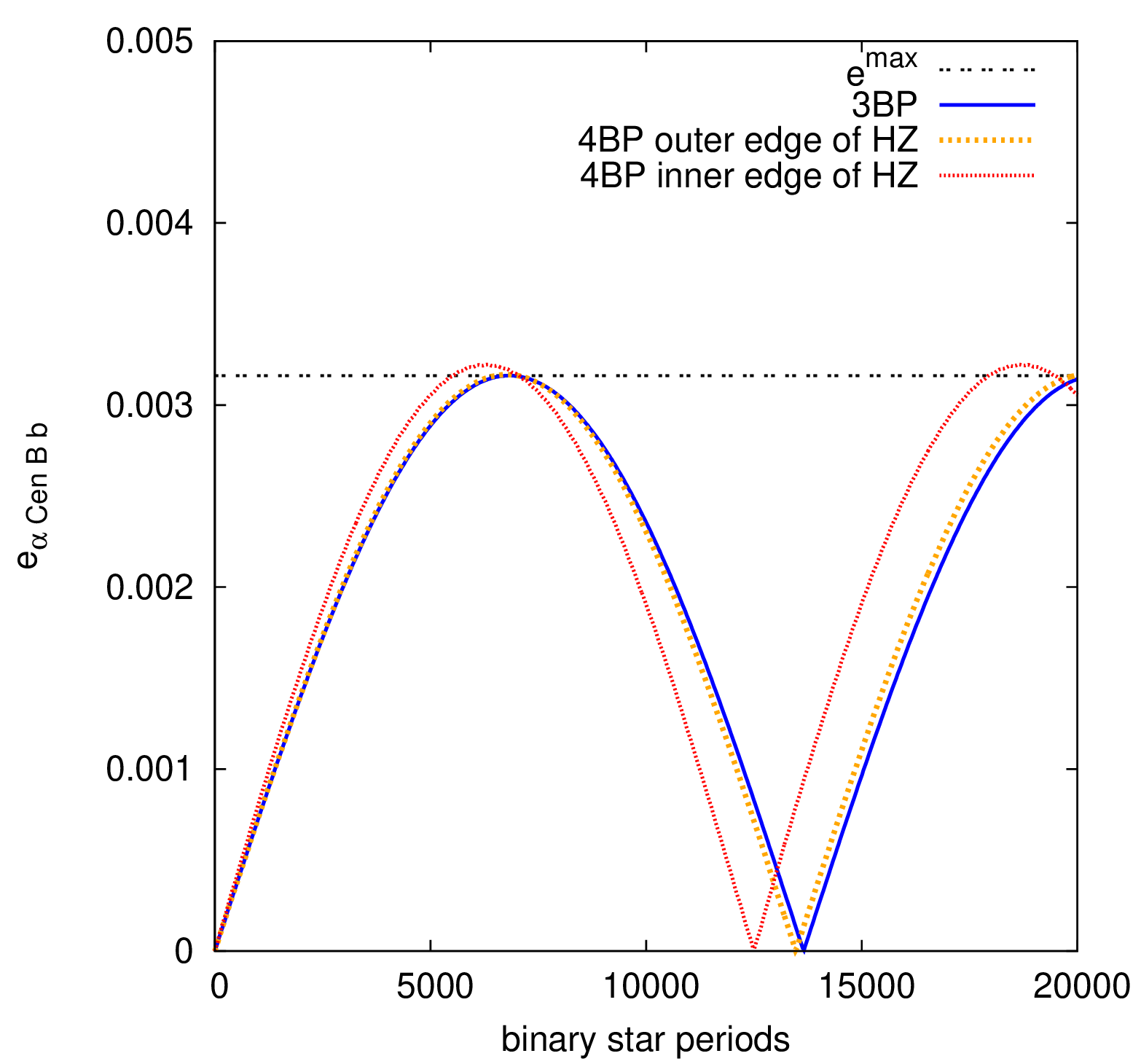} &
 \includegraphics[angle=-90, scale=0.34]{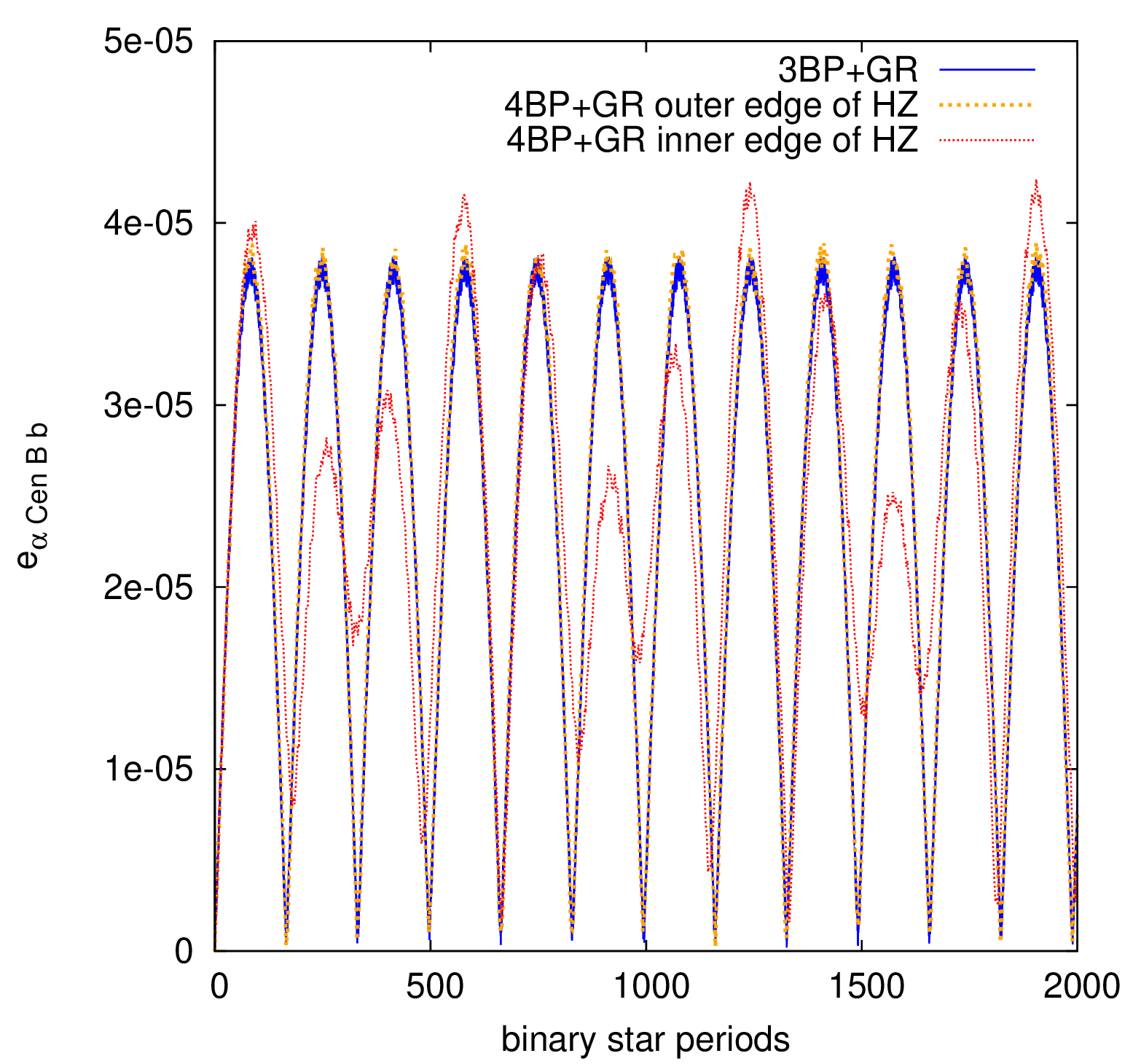}\\
 \includegraphics[angle=-90, scale=0.34]{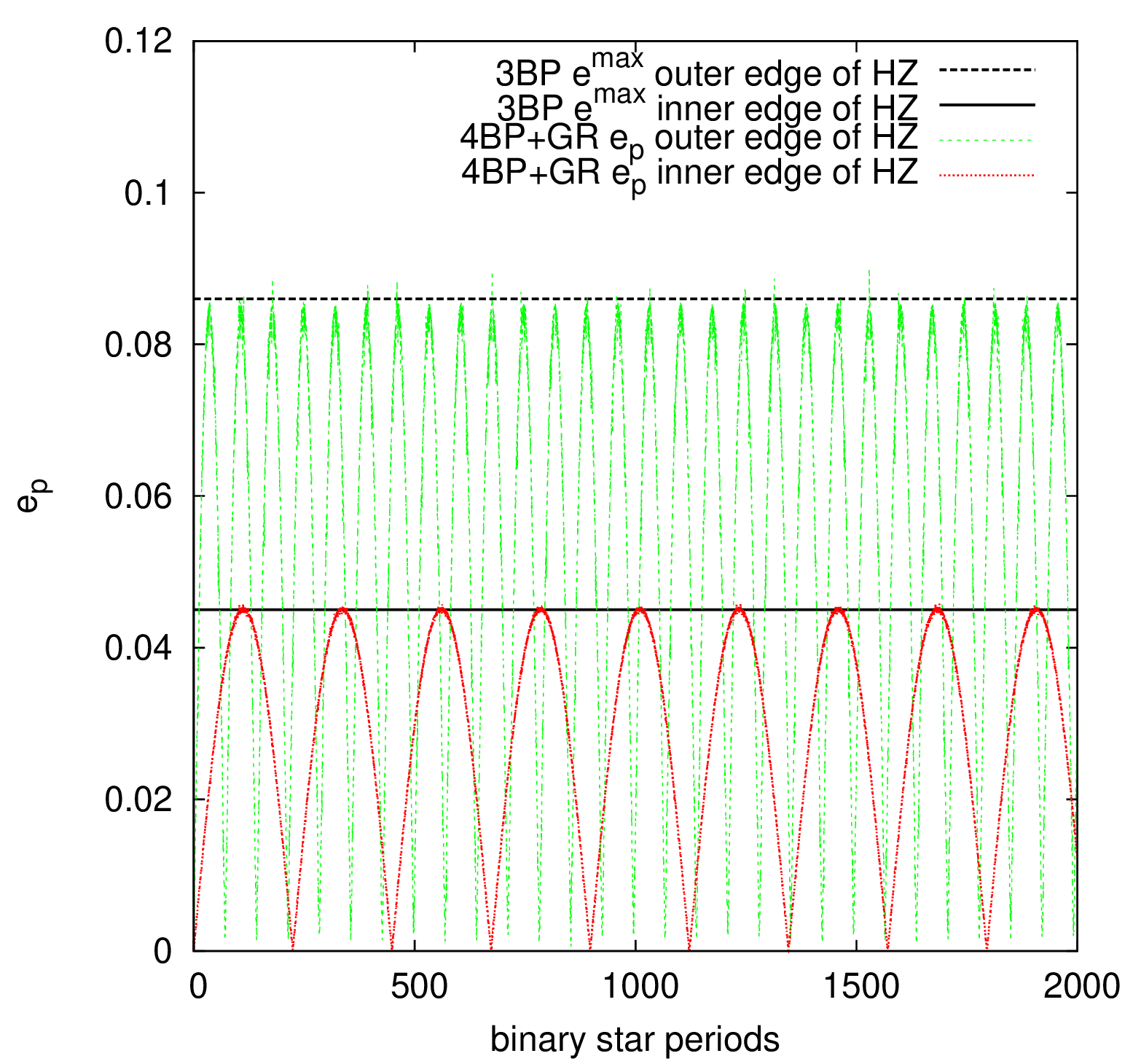}&
\includegraphics[angle=-90, scale=0.34]{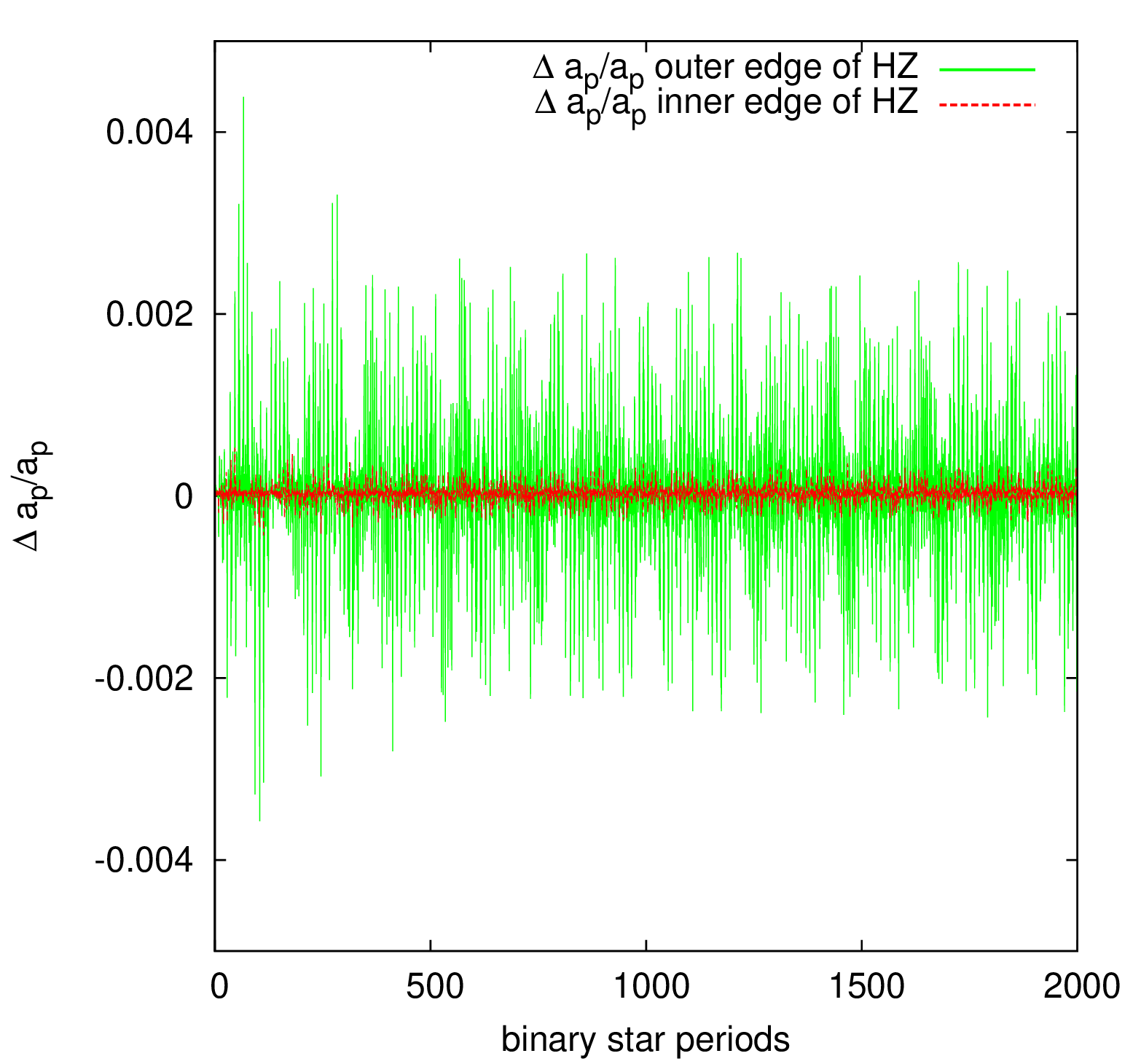}\\
\end{tabular}
\caption{An additional terrestrial planet in $\alpha$ Cen B's HZ affects the orbit of $\alpha$ Cen Bb (\textit{top row}) and vice versa (\textit{bottom row}).
%The influence of general relativity (GR) on the planets' eccentricities and semimajor axes are presented as well. 
In the top left panel, the numerically computed evolution of the eccentricity of $\alpha$ Cen Bb 
in the Newtonian three body problem (3BP) consiting of the binary $\alpha$ Cen AB and the planet $\alpha$ Cen Bb (blue) is compared with different four body problem scenarios (4BP). 
In one scenario, an additional Earth-sized body orbits $\alpha$ Cen B at the inner edge of its AHZ (red, see Table \ref{tab3}). In the other scenario the terrestrial planet is assumed to be at the outer edge of 
$\alpha$ Cen B's AHZ (orange). The analytic estimate for the maximum eccentricity ($e^{max}$) in the 3BP is presented as well (dotted, horizontal line).
The top right panel shows the exact same setup, only with general relativity (GR) taken into account. The orbit of $\alpha$ Cen Bb becomes practically circular.
While the influence of an additional planet at the outer edge of the HZ is barely noticable in the eccentricity evolution of $\alpha$ Cen Bb (blue, orange), a planet at the 
inner edge of the AHZ would cause distinct features (red).   
In contrast, neither GR nor $\alpha$ Cen Bb will influence the eccentricity evolution of planets ($e_p$) in the HZ significantly as is shown in the bottom left panel. Also, the semimajor
axes evolution of additional planets in the HZ is negligible (bottom right panel). Here, $\Delta a_p/a_p$ denotes the normalized difference between the 4BP+GR and the 3BP semimajor axis evolution of planets 
at the inner (red) and outer (green) edges of $\alpha$ Cen B's AHZ respectively.  
\label{fig6}}
\end{figure}

In figure \ref{fig6}, results of numerical investigations on the coupled orbital evolution of an additional terrestrial planet and $\alpha$ Cen Bb are presented. 
The top row of figure \ref{fig6} shows the eccentricity evolution of $\alpha$ Cen Bb altered by an additional Earth-like planet at the inner (red curve) and outer (orange curve) edge of $\alpha$ Cen B's AHZ .
The corresponding reference curve (blue) represents $\alpha$ Cen Bb's eccentricity influenced only by the binary $\alpha$ Cen AB.
The top left panel of figure 3 shows the results in Newtonian three (3BP) and four (4BP) body problems.
The top right panel depicts similar analysis with general relativity (GR) included.
The difference between the two approaches is quite pronounced, as GR clearly prevents the secular rise in Cen Bb's eccentricity predicted in the classical setup
\citep{blaes-et-al-2002,fabrycky-tremaine-2007}.
Thus, the orbit of $\alpha$ Cen Bb stays circular, even when tidal forces are neglected. 
The variations in semimajor axis ($\Delta a$) for $\alpha$ Cen Bb are not shown, because they remain below $10^{-8}$ AU for all cases.

A possible method to search for additional companions is to measure variations in $\alpha$ Cen Bb's orbital period. 
Yet, the small $\Delta a$ values make this approach difficult, since $\Delta P_p\propto P_p^{1/3} \Delta a$.
Disentangling the effects of GR and perturbations due to other habitable
planets on $\alpha$ Cen Bb's period would require precisions several orders of magnitude greater than
currently available.
%Changes in period due to an additional, habitable planet roughly 2-3 magnitudes smaller than the uncertainty given in dumusque et al.
The top right panel in figure \ref{fig6} shows that the perturbations an additional planet at the inner edge of $\alpha$ Cen B's AHZ causes in $\alpha$ Cen Bb's eccentricity (red) are, 
in principle, distinguishable from the nominal signal (blue). Unfortunately, 
it is also clear from this graph that neither the required precision nor the observational timescales necessary to identify the presence of an additional Earth-sized companion 
via observations of $\alpha$ Cen Bb's eccentricity 
seem obtainable in the near future.   
For habitable planets at the outer edge of $\alpha$ Cen B's AHZ the chances for indirect detection seem even worse, 
as their influence on $\alpha$ Cen Bb's orbit 
is negligible (orange vs. blue curves).  

In order to confirm that the interaction between $\alpha$ Cen Bb and Earth-like planets in the HZ is small, as well as to further study the 
influence of the GR on the dynamics of the system, we examined the orbital evolution of a fictitious habitable planet
in that region. The results are shown in the bottom row of figure \ref{fig6}.
The left panel depicts the eccentricity evolution of additional terrestrial planets positioned at the inner (red) and outer (green) edges of $\alpha$ Cen B's AHZ.
The secular variations in the eccentricity (bottom left panel) and semimajor axis (bottom right panel) of the habitable planet were computed numerically, taking the 
influence of the binary $\alpha$ Cen AB, the planet $\alpha$ Cen Bb, as well as GR into account. 
When comparing the analytic estimates of $e^{max}$ with the evolution of the habitable planet's eccentricity in the full system, 
it is evident that neither GR nor $\alpha$ Cen Bb alter the 
results for planets in $\alpha$ Cen B's HZ significantly. 
Also, the deviation in the habitable planet's semimajor axis due to GR and $\alpha$ Cen Bb ($\Delta a_p$) remains
below 0.1\% and 0.5\% for planets at the inner and outer edge of $\alpha$ Cen B's AHZ, respectively.
 
We conclude that the interaction between additional terrestrial planets in $\alpha$ Cen B's HZ and $\alpha$ Cen Bb is indeed small.
Thus, our estimates for the HZs of the $\alpha$ Centauri system remain valid.
The existence of additional terrestrial planets on the other hand cannot be determined easily from observing the orbital evolution of $\alpha$ Cen Bb.

The presented results are, strictly speaking, only valid for a coplanar configuration, i.e. the binary and both planets are in the same orbital plane. 
Mutually inclined configurations can exhibit much more involved dynamics such as Kozai resonant behavior (see e.g. \citet{correia-et-al-2011}). 
A detailed study of such effects lies beyond the scope of this work. Nevertheless, the arguments presented in this section suggest that the search for an additional coplanar planet in the HZ around $\alpha$ Cen B will most likely have to be performed without relying on 
observations of $\alpha$ Cen Bb.
We will, therefore, investigate whether habitable planets can actually be detected independently in Sun-like binary star configurations using current observational facilities.

\subsection{Detectability Through Radial Velocity and Astrometry}\label{sec:detect}
We apply our methodology, as derived in sections \ref{sec:rv} and \ref{sec:astrometry}, to a fictitious 
terrestrial planet in the HZ of binary systems similar to $\alpha$ Centauri AB but with a broadened range of binary eccentricities. 
In addition to the habitability maps discussed in section \ref{sec:hz}, figures \ref{fig3} and \ref{fig4} show the results regarding peak and RMS strength 
of the RV and astrometric signals. Here, the aim is to illustrate how the different types of HZs presented in section 
\ref{sec:hz}, as well as the maximum and RMS signal strengths defined in section \ref{sec:rv} vary with 
the binary's eccentricity ($e_b$) and planetary semimajor axis ($a_p$). 
The left column of figure \ref{fig3} shows maximum (\textit{top}) and RMS  (\textit{bottom}) values of the signal strengths for the more massive binary component, in this case similar to $\alpha$ Cen A. 
Results for the less massive component akin to $\alpha$ Cen B are shown in the right column. 
%The horizontal line indicates $\alpha$ Centauri's actual eccentricity $e_b=0.5179$. 

The dashed vertical lines in the top rows of figures \ref{fig3} and \ref{fig4} represent the sections of the parameter space 
with similar $V_r^{circ}$ and $\rho^{circ}$ values, respectively. Since $V_r^{circ}$ and $\rho^{circ}$ are independent of the planetary (and consequently the binary's) 
eccentricity, the different values of these quantities vary linearly with the planet's semimajor axis. 
In contrast, $V_r^{max}$ and $\rho^{max}$, represented by the solid contour lines, depend on the maximum eccentricity of the planet ($e_p^{max}$) and therefore change
with the binary's eccentricity ($e_b$). Since for circular binary configurations only small eccentricities are induced into the planet's orbit, 
$V_r^{max}$ and $V_r^{circ}$ almost coincide. The same holds true for $\rho^{max}$ and $\rho^{circ}$ in this case. 
Yet, $V_r^{max}$ and $\rho^{max}$ grow with the binary's eccentricity.
The corresponding contour lines indicate that for high binary eccentricities 
even small planetary semimajor axes can produce similar AM peak signal strength.
Similarly, planets with larger distances to their host stars can still cause similar RV amplitudes if the binary's eccentricity is sufficiently large.  
If a fixed detection limit is set, e.g. $V_r = 9.5$ m/s, planets with semimajor axes up 
to $1.5$ AU could still be found around stars similar to $\alpha$ Cen A, assuming a binary eccentricity of $e_b=0.7$.
To produce a similarly high RV amplitude, a circular planet has to orbit its host star at roughly $0.8$ AU (Fig.~\ref{fig3}).  
In other words, high binary eccentricities lead to excited planetary eccentricities which in turn increase the peak signal strengths suggesting that
binary-planet interactions can actually improve the chances for detecting terrestrial planets.
Naturally, if the planet's eccentricity happens to be close to zero at the time of observation, this advantage is nullified.

The bottom row of Fig.~\ref{fig3} shows the same setup with RMS signal strengths $\langle\langle V_r \rangle\rangle_{M,\omega}$ and $\langle\langle \rho \rangle\rangle_{M,\omega}$,
respectively.
While $\langle\langle V_r \rangle\rangle_{M,\omega}$ is independent of the binary's eccentricity,
it is evident from equation (\ref{eq:rhorms}) that  
 $\langle\langle \rho \rangle\rangle_{M,\omega}$ depends weakly on $e_b$ since $\langle e_p^2 \rangle ^{1/2}\simeq 0.1$ 
for the cases considered and therefore $\langle e_p^2\rangle \ll 1$ (see Fig.~\ref{fig5}, \textit{bottom}).
The slight curvature of the contour lines representing the RMS signal in Fig.~\ref{fig4} indicates this behavior.
A summery of RV and AM signal strengths for an Earth-like planet at the boundaries of $\alpha$ Centauri's HZ is presented in Table~\ref{tab3}.

We illustrated in this section that the dynamical interactions between a terrestrial planet and the secondary
star can produce large peak amplitudes which may enhance the detectability of the planet with the RV
and AM methods considerably. The RMS values of the planet's AM and RV signals, on the other hand, remain almost unaffected by the 
gravitational influence of the secondary star.

\subsection{Transit Photometry}

\begin{figure}
\center
\includegraphics[angle=-90, scale=0.40]{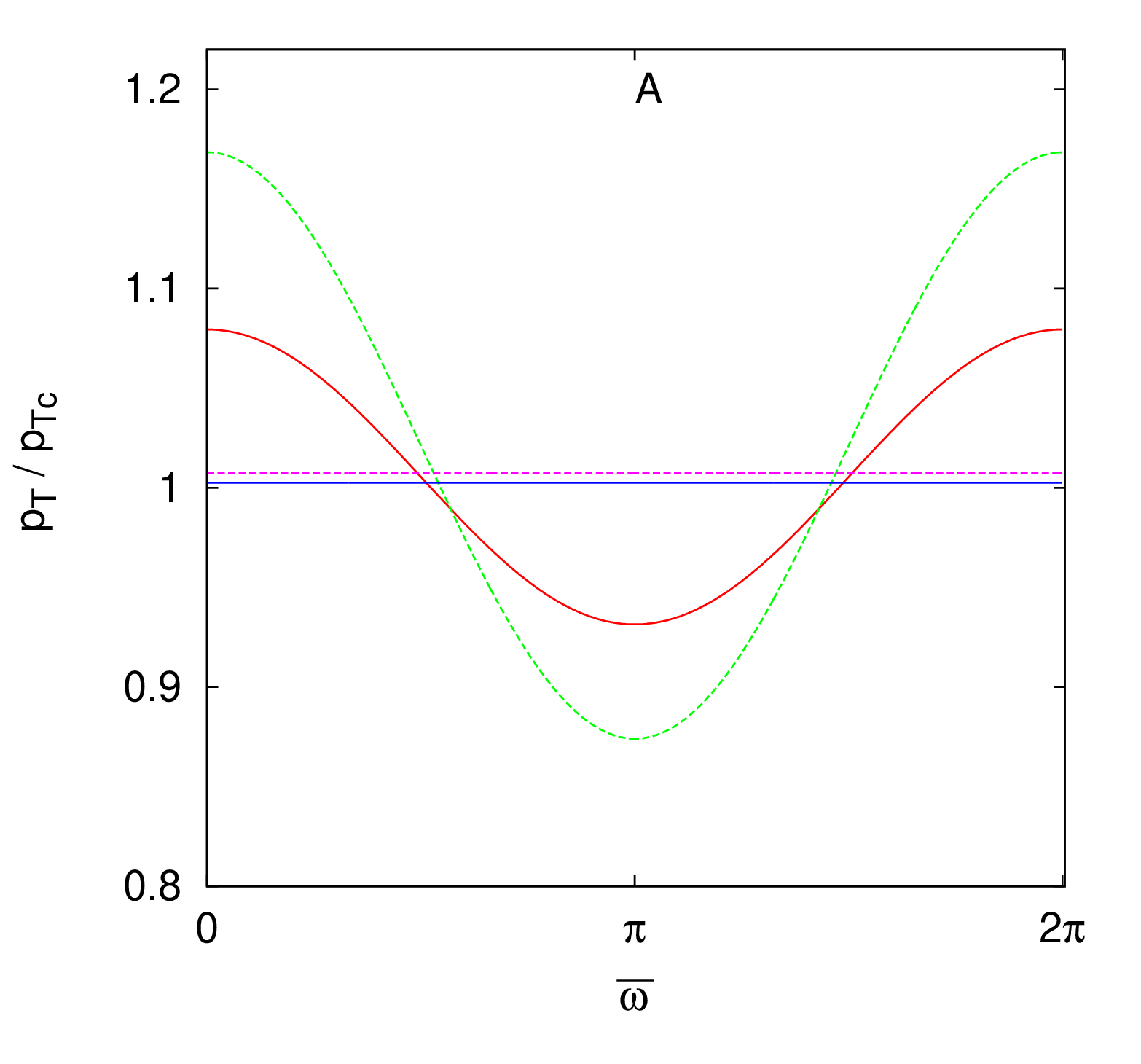} 
\includegraphics[angle=-90, scale=0.40]{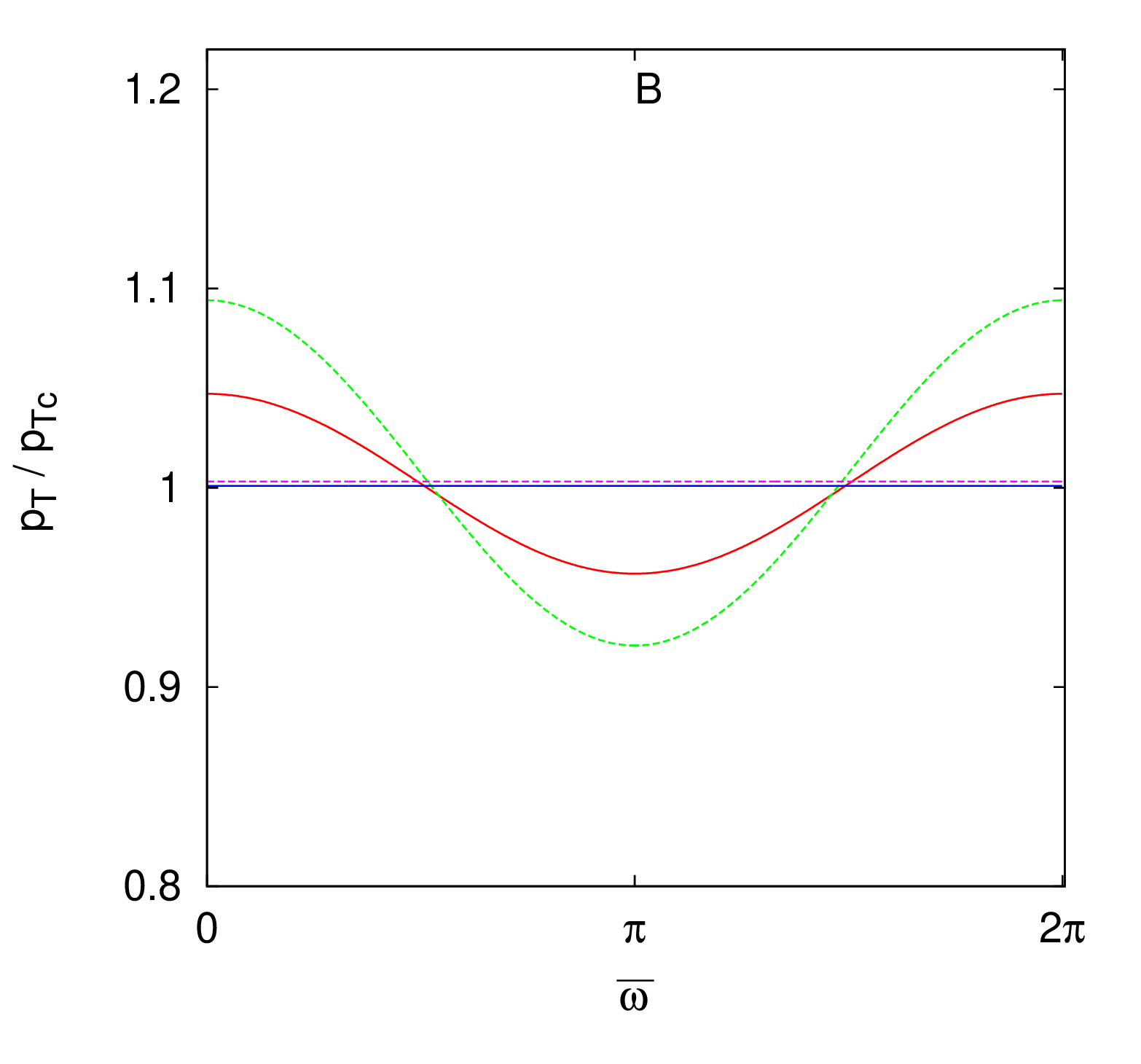} 
\caption{Graphs of the ratio of transit probabilities ($p_T/p_{Tc}$) with ${p_T}\equiv{p_T}|_{e=e^{max}}$
and ${p_{Tc}}\equiv{p_T}|_{e=0}$, in a binary similar to the $\alpha$ Centauri system. The graphs show the
transit probabilities in terms of the planet's argument of pericenter ($\bar{\omega}$), as measured from the 
line of sight for component A (top) and component B (bottom). The red, full line corresponds to the planet starting 
at the inner border of AHZ and dashed-green corresponds to the outer edge, for each star (see Table~\ref{tab3}). 
The full and dashed straight lines are the corresponding ratios of the averaged transit probabilities 
($\langle p_T \rangle /  \langle p_{Tc}\rangle$) evaluated using equation (\ref{eq:prob4}). 
Compared to the transit probability ratios $p_T/p_{Tc}$, the ratio of averages
$\langle p_T \rangle /  \langle p_{Tc}\rangle$  shows only a weak dependence on the planet's 
initial position in the HZ. 
\label{fig7}} 
\end{figure}

To assess the detectability of a terrestrial planet in the HZ of $\alpha$ Centauri AB (and similar
binaries) through transit photometry, we calculated the relative transit depths that an Earth-like 
planet would produce during its transit. If such a system hosted a transiting terrestrial planet, 
TD values would range around 55 ppm for $\alpha$ Cen A, and 115 ppm for  $\alpha$ Cen B. Such transit 
depths are detectable by NASA's \textit{Kepler} telescope for instance -- stellar and instrumental 
sources included -- as the spacecraft's median noise level amounts to $\approx$ 29 ppm 
\citep{gilliland-et-al-2011}. Therefore, Earth-like planets could in theory be found around $\alpha$ 
Centauri stars. However, \textit{Kepler} was not designed to observe stars with apparent magnitudes between 
0 and 3 such as those of $\alpha$ Centauri. 
The TESS mission (Transiting Exoplanets Survey Satellite), for instance, will aim for TP of brighter stars \citep{ricker-et-al-2010}.
Nevertheless, the example of \textit{Kepler} suggests that the
detection of  transiting habitable planets in S-type systems would be possible using current technology. 
In fact, very much similar to the cases discussed in the previous sections, the eccentricity excitation 
that an Earth-like planet experiences in a binary star system may enhance its possibility of detection 
via transit photometry \citep[][also see Fig.~\ref{fig7}]{Kane08,Kane12,brokovits-et-al-2003}. Assuming  $\alpha$ Centauri 
was a transiting 
system\footnote{$i_b=90^\circ\pm\theta_{planet}/2$ \citep{borucki-summers-1984}.}, a comparison 
of the transit probabilities of actual planetary orbits to circular orbits shows that an 18\% increase 
in $p_T$ values seems possible for terrestrial planets at the outer edge of  $\alpha$ Cen A's AHZ 
(Fig.~\ref{fig7}). Given the right orbital configuration, it may be more likely to identify a 
transiting habitable terrestrial planet around a stellar component of a binary than around a single star assuming similar initial
planetary eccentricities. 
 
The increase in transit probability for planets in double star systems is less dramatic when the 
equations are averaged over all possible configurations of the argument of pericenter as in equation 
(\ref{eq:prob4}). Averaged transit probabilities are represented by the straight lines in figure \ref{fig7}. 
As $\langle p_T \rangle /  \langle p_{T}|_{e=0} \rangle > 1$, for terrestrial planets' orbits with $e>0$, 
the chance for transit is in general higher for Earth-like planets in binary stars 
than for terrestrial planets in circular orbits around single stars.

\section{Discussion}\label{sec:dis} 

Comparing the quantitative estimates of RV, AM and TP signals, TP seems to be the best choice for finding 
Earth-like planets in the HZs of a coplanar S-Type binary configuration with Sun-like components. 
Even for a system as near as $\alpha$ 
Centauri, AM peak signals only measure $\mu$as.  
Unfortunately, neither ESO's VLBI with PRIMA, nor ESA's GAIA mission will be able to deliver such precision 
in the near future \citep{quirrenbach-et-al-2011}. GAIA's aim to provide $\mu$as astrometry will most 
likely not be achieved until the end of the mission \citep{hestroffer-et-al-2010}. Also, from an astrometric 
point of view, Earth-like planets would be easier to find around  $\alpha$ Cen A  than $\alpha$ Cen B. 
That is because the HZ around star A is more distant from this star. Naturally, the opposite is true 
for RV detections. Due to the difference in the stellar masses, $\alpha$ Cen B offers a better chance of 
finding a terrestrial planet there using RV techniques. The recent discovery of an Earth-sized planet around
this star supports our results. The observed planetary RV signal was reproduced excellently by our analytic estimates for circular planetary orbits. 

Our prediction of RV amplitudes for terrestrial planets around $\alpha$ Cen B are also in good agreement 
with those presented by \citet{guedes-et-al-2008}.   
The four terrestrial planets used in the RV model
by these authors produce almost exactly four times the predicted RMS amplitude given in figure \ref{fig3}.
\citet{guedes-et-al-2008} claim that Earth-like planets in the $\alpha$ Centauri are detectable even for 
signal-to-noise ratios (SNR) of single observations below 0.1. However, obtaining sufficient data to reconstruct the planetary signal
requires a great amount of dedicated observing time (approximately 5 years in their example).
Validating this statement, it took \citet{dumusque2012} about 4 years of acquired data to detect $\alpha$ Cen Bb. 
The data published by \citet{dumusque2012} also allow a glimpse on the current performance of the HARPS spectrograph
revealing a precision around $50-80$ cm/s. 
Given the fact that the RV signal of a habitable planet around $\alpha$ Cen B would be still half an order of magnitude smaller (Fig.~\ref{fig3}),
considerably more observation time would be required to identify habitable companions. 
HIRES measurements are currently yielding precisions around $1$ m/s. Identifying RV signals of habitable worlds around 
$\alpha$ Cen B therefore seems even more unlikely when using HIRES.  
The previous examples show that some development of observational capacities is still necessary to achieve the RV resolution required for discovering
habitable planets in the $\alpha$ Centauri system. 

The success of NASA's \textit{Kepler} space telescope in identifying countless Earth-sized 
planetary candidates \citep[e.g.,][]{kepler2-2011} that require follow-up observations might provide 
the necessary momentum to develop instruments capable of resolving RV signals in the range of cm/s. 
Focusing on less massive binaries would have the advantage of having greatly enhanced RV signals 
as the HZs will be situated closer to the planet's host stars. How far this might simplify the 
task of finding habitable worlds will be the topic of further investigations.

In regard to transit photometry, both \textit{Kepler} and CoRoT telescopes have proven that it is possible 
to find terrestrial planets around Sun-like stars \citep[e.g.,][]{leger-et-al-2009,borucki-et-al-2012}. 
The combination of proven technology and the presented argument that the dynamical environment in 
binary star systems will enhance transit probabilities makes photometry currently the most promising 
method for finding Earth-like planets in the HZs of S-Type binary star systems.

\section{Summary}

In this work, we provided an analytic framework to estimate the detectability of a terrestrial 
planet using radial velocity (RV), astrometry (AM), as well as transit photometry (TP) in coplanar 
S-Type binary configurations. We have shown that the gravitational interactions between the stars 
of a binary and a terrestrial planet can improve the chances for the planet's detection. The induced 
changes in the planet's eccentricity enhance not only RV and AM peak amplitudes, but also the 
probability to witness a planetary transit. Next to the presented "best case" estimates, we 
offered RMS/averaged expressions which are deemed to be more suited to determine the long-term 
influence of the second star on planetary fingerprints in S-Type systems. In contrast to peak 
amplitudes, the RMS of a planet's AM signal is only modified slightly by the additional gravitational 
interaction with the second star. A similar behavior can be seen in planetary transit probabilities.
The RMS values of RV signals are altogether independent of the secondary's gravitational influence, assuming that 
the system is nearly coplanar.

After defining the Permanent, Extended, and Average Habitable Zones for both stellar components 
of the $\alpha$ Centauri system, we investigated the possible interaction between the newly discovered $\alpha$ Cen Bb
and additional terrestrial companions in $\alpha$ Cen B's HZ. Our results suggest 
that $\alpha$ Cen Bb is on an orbit with very low eccentricity which would not be influenced significantly by habitable, terrestrial
companions. Conversely, $\alpha$ Cen Bb's presence would also not affect Earth-like planets in the habitable zone of $\alpha$ Cen B.
   
We estimated the maximum and RMS values of the RV as well as AM signal 
for a terrestrial planet in the $\alpha$ Centauri habitable zones. The peak and RMS amplitudes of the RV signal 
ranged between $4$ and $12$ cm/s. Astrometric signals were estimated to lie between $1$ and 5 $\mu$as.
Given the current observational facilities, enormous amounts of observing time would be required 
to achieve such precisions. If the $\alpha$ Centauri was a transiting system, however, a habitable planet 
could be detectable using current technologies. It seems that the detection of Earth-like planets 
in circumstellar habitable zones of binaries with Sun-like components via astrometry and radial 
velocity is still somewhat beyond our grasp, leaving photometry to be the only current option in this respect.

\acknowledgments

SE and EP-L acknowledge support from FWF through projects AS11608-N16 (EP-L and SE), 
P20216-N16 (SE and EP-L), and P22603-N16 (EP-L). SE acknowledges support from the University of Vienna's 
Forschungsstipendium 2012. NH acknowledges support from the NASA Astrobiology Institute under Cooperative 
Agreement NNA09DA77A at the Institute for Astronomy, University of Hawaii, and NASA EXOB grant NNX09AN05G. 
SE and EP-L would also like to thank the Institute for Astronomy and NASA Astrobiology Institute at the University 
of Hawaii-Manoa for their kind hospitality during the course of this project. The authors are thankful to 
Nikolaos Georgakarakos for his valuable suggestions and to the anonymous referee for his/her constructive
comments.

\onecolumn

\appendix 

\section{Equation of the Center}\label{sec:eoc}
The equation of the center providing a direct relation between the true anomaly $f$ and the mean anomaly $M$ is presented up to the $6^{th}$ order in eccentricity $e$:
\begin{eqnarray}
f=M+\left(2 e-\frac{e^3}{4}+\frac{5 e^5}{96}\right) \sin{M}+ \left(\frac{5 e^2}{4}-\frac{11 e^4}{24}+\frac{17 e^6}{192}\right) \sin[2 M] \\ \nonumber+\left(\frac{13 e^3}{12}-\frac{43 e^5}{64}\right) \sin[3 M] +\left(\frac{103 e^4}{96}-\frac{451 e^6}{480}\right) \sin[4 M] \\ \nonumber+\frac{1097}{960} e^5 \sin[5 M] +\frac{1223}{960} e^6 \sin[6 M] + O(e^7)
\end{eqnarray}

\section{Averaging of $\rho^2$} \label{sec:ave2}

The averaging integrations over $M$ and $\omega$ in equations (\ref{eq:rhorms}) and (\ref{eq:rhorms2}) 
were carried out as in the following;

\begin{displaymath}
\frac{1}{4\pi^2}\iint\limits_0^{\quad 2\pi} \rho^2(M,\omega) dM d\omega  = 
\frac{1}{2\pi}\int_0^{2\pi}\frac{\mu^2 a^2}{d^2} 
\left\{1+\frac{3 \langle e^2 \rangle_M}{2}+ \left[-\frac{1}{2}+\frac{\langle e^2\rangle_M}{4} 
(5 \cos(2 \omega )-3)\right] 
\sin^2{i}\right\}d\omega \,.
\end{displaymath}

\noindent
The integration over $M$ is trivial. Using  the partial integration technique to integrate over $\omega$, 
we obtain

 \begin{displaymath}
\frac{1}{2\pi}\int\limits_0^{2\pi} \frac{5}{4}\langle e^2\rangle_M  \cos(2 \omega) d\omega  = 
\frac{5}{4} \left[\langle e^2\rangle_{M,\omega} \cos(2 \omega)|_0^{2\pi}+ 
2\langle e^2\rangle_{M,\omega}\int_0^{2\pi}\sin(2\omega)d\omega \right] =0\\ \nonumber
\end{displaymath}

\noindent
Here we have used the fact that 
$\langle e^2\rangle_{M,\omega}=(1/2\pi)\int\limits_0^{2\pi}\langle e^2\rangle_{M}d\omega$ 
does no longer depend on $\omega$. From the definition of averaging given by equation (\ref{eq:rms}), we have

\begin{eqnarray}
 \langle\langle \rho \rangle\rangle_{M,\omega} &=&\frac{\mu a}{2d}\left[3 + 
\frac{9}{2} \langle e^2 \rangle_{M,\omega}  + 
\left(1 + \frac{3}{2} \langle e^2 \rangle_{M,\omega} \right) \cos(2 i)\right]^{1/2}
\end{eqnarray}

\noindent
A similar procedure has been applied to derive equation (\ref{eq:vrrms}).

\clearpage
% 
% \begin{figure}
% \center
% \includegraphics[angle=-90, scale=0.45]{fig1a.eps} 
% \includegraphics[angle=-90, scale=0.45]{fig1b.eps} 
% \caption{
% % Graph of the variations of the planet's orbital eccentricity 
% % in a G2V-G2V S-Type binary system. The semimajor axis of the binary is
% % 20 AU and its eccentricity was varied from 0.1 to 0.7. The planet was initially
% % in a circular orbit at 1 AU around its host star. As shown here, even though 
% % the planet started in a circular orbit, the perturbation of the secondary star
% % caused its orbital eccentricity to increase.
% This figure shows the correlation between the evolution of planetary eccentricity and insolation in a G2V-G2V \textit{S-Type} planet-binary configuration.
% The planet was started on an initially circular orbit ($a_p= 1$AU) around one of the two Sun-like stars ($a_b=20$AU).
% The left panel shows that even though the planet was started on a circular orbit, the perturbation of the secondary star
% caused its orbital eccentricity to vary periodically.
% The corresponding amplitude variations in planetary insolation are depicted in the right panel. 
% Incident flux values ($S_{eff}$) are given in units of Solar constants.
% \label{fig1}} 
% \end{figure}
% 
% 

\clearpage

\clearpage

\clearpage

\clearpage

\clearpage

\clearpage

\clearpage

\begin{deluxetable}{l|cc}
%\tabletypesize{\scriptsize}
%\rotate
\tablecaption{Physical and orbital parameters of the $\alpha$ Centauri AB\textsl{b} system 
\citep{kervella-et-al-2003,guedes-et-al-2008,pourbaix-et-al-2002,dumusque2012}. 
\label{tab1}}
\tablewidth{0pt}
\tablehead{\colhead{$\alpha$ Centauri} & \colhead{A} &\colhead{B}}
\startdata
spectral classification & G2V & K1V\\
mass [$M_\odot$] & 1.105 $\pm$ 0.007 & 0.934 $\pm$ 0.007\\
$T_{eff}$ [K] & 5790 & 5260\\
luminosity [$L_\odot$] & 1.519 & 0.500\\\hline
distance [pc] & \multicolumn{2}{c}{1.339 $\pm$ 0.002} \\
period ($P_b$) [d] & \multicolumn{2}{c}{29187 $\pm$ 4} \\
$a_b$ [AU] & \multicolumn{2}{c}{23.4 $\pm$ 0.03}\\
$e_b$ & \multicolumn{2}{c}{0.5179 $\pm$ 0.00076}\\
$i_b$ [deg] & \multicolumn{2}{c}{79.205 $\pm$ 0.0041 }\\
$\omega_b$ [deg]  & \multicolumn{2}{c}{231.65 $\pm$ 0.076 }\\
$\Omega_b$ [deg]  &\multicolumn{2}{c}{204.85 $\pm$ 0.084 }\\\hline
\colhead{$\alpha$ Centauri} & \multicolumn{2}{c}{B b}\\\hline
$P_p$ [d] & \multicolumn{2}{c}{3.2357 $\pm$0.0008}\\
$e_p$ &  \multicolumn{2}{c}{0 (fixed)}\\
minimum mass ($m_p^{min}$) [$M_{Earth}$]& \multicolumn{2}{c}{1.13 $\pm$0.09}
\enddata
\end{deluxetable}

\begin{deluxetable}{lcc}
\tablecaption{A comparison between the predicted RV values using the analytic expressions derived in section \ref{sec:rv} 
and the observed values for the terrestrial planet discovered around $\alpha$ Cen B \citep{dumusque2012}.
Coplanarity of the system was assumed.
The formal uncertainties have been derived assuming Gaussian error propagation of the uncertainties given in Tab.~\ref{tab1}. The maximum 
predicted planetary eccentricity for $\alpha$ Cen B\textsl{b} is $e^{max}=0.003$.
When taking general relativity into account, however, the orbit of $\alpha$ Cen B\textsl{b} will remain practically circular (see Fig.~\ref{fig6}).\label{tab2}} 
\tablehead{\colhead{}& \colhead{Predicted signal [m/s]} & \colhead{Observed signal [m/s]}}
\tablewidth{0pt}
\startdata
$\langle\langle V_r \rangle\rangle_{M,\omega}$ & 0.365 $\pm$0.029 &\\ 
$V_r^{circ}$                                   & 0.517 $\pm$0.041                          &     0.51 $\pm$ 0.04\\
$V_r^{max}$                                    & 0.519 $\pm$0.041 &\\                                
\enddata
\end{deluxetable}

\clearpage

\begin{deluxetable}{c|c|cccccc|cl}
\tabletypesize{\scriptsize}
%\rotate
\tablecaption{Detectability of an Earth-like planet in the HZs of the $\alpha$ Centauri system. 
Each row shows the maximum amplitude of the radial velocity signal as well as the astrometric fingerprints of a terrestrial 
planet in the $\alpha$ Centauri HZs. The critical planetary semimajor axis ($a_c$) indicates the onset of dynamical instability 
\citep{holman-wiegert-1999}. Computations using chaos indicators are in good agreement with those stability limits 
\citep{pilat-lohinger-dvorak-2002}. Analytic expressions for calculating the boundary values of planetary semimajor axes in the system's HZs are 
given in \citep{eggl-et-al-2012}.
\label{tab3}}
\tablewidth{0pt}
\tablehead{  $\alpha$ Cen & $a_{c}$ [AU]&  inner AHZ &  inner EHZ &  inner PHZ  & outer PHZ & outer EHZ  & outer AHZ & }
\startdata

  \multirow{5}{*}{A} &     \multirow{5}{*}{2.76}  &          1.03 &          1.07 &          1.12 &          1.81 &          1.94 &          2.06& HZ border & [AU] \\\cline{3-10}
  & &          8.97 &          8.83 &          8.66 &          7.14 &          6.97 &          6.82 & $V_r^{max}$                                     & \multirow{2}{*}{[cm/s]} \\
  & &          5.89 &          5.78 &          5.65 &          4.44 &          4.30 &          4.17 & $\langle\langle V_r\rangle\rangle_{M,\omega}$   & \\\cline{3-10}
  & &          2.28 &          2.37 &          2.49 &          4.20 &          4.52 &          4.84 & $\rho^{max}$                                    & \multirow{2}{*}{[$\mu$as]} \\
  & &           1.53 &          1.59 &          1.66 &          2.69 &          2.88 &          3.06 & $\langle\langle \rho \rangle\rangle_{M,\omega}$ & \\\hline
   \multirow{5}{*}{B} &      \multirow{5}{*}{2.51} &          0.62 &          0.64 &          0.65 &          1.13 &          1.19 &          1.23    & HZ border & [AU]  \\\cline{3-10}
  & &          12.21 &         12.09 &         11.94 &          9.37 &          9.19 &          9.04 & $V_r^{max}$                                     & \multirow{2}{*}{[cm/s]}\\
  & &          8.25 &          8.16 &          8.05 &          6.12 &          5.98 &          5.86 & $\langle\langle V_r\rangle\rangle_{M,\omega}$   & \\\cline{3-10}
  & &          1.58 &          1.62 &          1.66 &          2.97 &          3.12 &          3.26 & $\rho^{max}$                                    & \multirow{2}{*}{[$\mu$as]} \\
  & &          1.09 &          1.11 &          1.14 &          1.98 &          2.08 &          2.16 & $\langle\langle \rho \rangle\rangle_{M,\omega}$ &\\\hline
\enddata
\end{deluxetable}

\end{document}